\documentclass[11pt,a4paper]{article}
\usepackage[utf8]{inputenc}
\usepackage{jheppub}
\usepackage{mathtools}
\usepackage{graphicx}
\usepackage{booktabs}

\allowdisplaybreaks

\newcommand\identity{1\kern-0.25em\text{l}\kern+0.07em}

\def\Res_#1{\operatorname*{Res}_{#1}}

\def\Tr{\operatorname*{Tr}}

\def\nn{\nonumber}
\def\bra#1{\langle #1|}
\def\ket#1{|#1 \rangle}
\def\braket#1{\langle #1 \rangle}

\def\dd{{\rm d}}
\def\ddd{{\rm d}\!\!\!{}^-\;\!\!}
\def\del{\delta\!\!\!{}^-\;\!\!}

\def\ie{i.e. }

\def\eqn#1{eq.~\eqref{#1}}
\def\eqns#1#2{eqs.~\eqref{#1} and~\eqref{#2}}

\def\Eqn#1{Eq.~\eqref{#1}}

\def\tab#1{table~{\ref{#1}}}

\def\sec#1{section~{\ref{#1}}}

\def\app#1{appendix~{\ref{#1}}}

\def\foot#1{footnote~{\ref{#1}}}

\newcommand{\W}{{\rm W}}
\newcommand{\cev}[1]{\reflectbox{\ensuremath{\vec{\reflectbox{\ensuremath{#1}}}}}}
\title{Classical dynamics from QFT via Stratonovich-Weyl correspondence}
\author[1]{Alexander Ochirov}
\author[2]{and Canxin Shi}

\affiliation[1]{Center for Fundamental Physics, School of Physical Science and Technology, \\
ShanghaiTech University, 393 Middle Huaxia Road, Shanghai 201210, China}
\affiliation[2]{New Cornerstone Laboratory, Institute of Theoretical Physics, Chinese Academy of Sciences, Beijing 100190, China}

\emailAdd{ochirov@shanghaitech.edu.cn,shicanxin@itp.ac.cn}

\abstract{We develop a formalism for extracting relativistic classical dynamics from the S-matrix, in which massive bodies are described on phase space while massless radiation is kept in the Fock space. Our formalism relies on the Stratonovich-Weyl (SW) correspondence between the operator and phase-space formulations of quantum mechanics, as implemented covariantly by the SW quantizer. For localized incoming states, classical observables are given by SW symbols of the appropriate operators evaluated on incoming asymptotic trajectories. In combination with the exponential form of the S-matrix, the observables are naturally given by nested mixed Dirac/commutator brackets involving Magnus amplitudes. The latter are the matrix elements of the logarithm of the S-matrix, and we provide a straightforward algorithm for computing the combinatorial Murua coefficients that control the placement of retarded propagators inside Magnus-amplitude diagrams. This picture is well-suited for gravitational-wave physics applications.
}

\begin{document}
\maketitle
\addtocontents{toc}{\protect\setcounter{tocdepth}{2}}

\section{Introduction}
\label{sec:intro}

It is by now common knowledge that the methods of quantum field theory (QFT) may be successfully used to extract new interesting classical-gravity results~\cite{Guevara:2018wpp,Bern:2019nnu,Bern:2019crd}.
In this paradigm, the quantum-physics information is discarded in favor of the classically relevant contributions while preserving the advantages of scattering-amplitude-based calculations, notably gauge invariance.
The most widely used ways for this nontrivial transition are: the effective-field-theory (EFT) matching approach~\cite{Neill:2013wsa,Cheung:2018wkq,Bern:2019nnu,Bern:2019crd,Bern:2020buy,Bern:2021yeh,Bern:2026oqp}, the Kosower-Maybee-O'Connell (KMOC) formalism~\cite{Kosower:2018adc,Maybee:2019jus,delaCruz:2020bbn,Cristofoli:2021vyo,Aoude:2021oqj,Britto:2021pud,Cristofoli:2021jas} centered around QFT observables, the eikonal exponentiation~\cite{Kabat:1992tb,Akhoury:2013yua,Bjerrum-Bohr:2018xdl,KoemansCollado:2019ggb,Cristofoli:2020uzm,
AccettulliHuber:2020oou,Bjerrum-Bohr:2021vuf,
DiVecchia:2021bdo,Bjerrum-Bohr:2021din,DiVecchia:2023frv} in the impact-parameter space
and heavy-particle EFTs~\cite{Damgaard:2019lfh,Aoude:2020onz,Brandhuber:2021eyq,Brandhuber:2023hhy} analogous to heavy-quark effective theory~\cite{Georgi:1990um,Bodwin:1994jh}.\footnote{Of course, there are also major developments in the application of classical worldline methods~\cite{Driesse:2024feo,Brunello:2025gpf,Driesse:2026qiz,Porto:2026fsd,Dlapa:2026oyq,Brunello:2026anu}.
}

More recently, it has been realized that the KMOC formalism could be connected~\cite{Damgaard:2023vnx,Capatti:2024bid} to the classical worldline approach.
Moreover, the exponential form~\cite{Bjerrum-Bohr:2021wwt,Damgaard:2021ipf} of the S-matrix has been used to link the eikonal to both the KMOC formalism~\cite{Damgaard:2023ttc} and to the so-called radial action obtained from the four-point amplitude~\cite{Kim:2025gis}.
Such studies are interesting in the context of adapting the KMOC formalism, which is conceptually transparent but is inherently quantum-field-theoretic, to classical-physics applications and thus making it more computationally efficient.
One such hybrid framework that applies classical Poisson/Dirac brackets to scattering amplitudes has been identified by Alessio, Gonzo and one of the current authors~\cite{Gonzo:2024zxo,Alessio:2025flu} and successfully used for rotating black-hole scattering calculations by Akpinar et al.~\cite{Akpinar:2025bkt,Akpinar:2025tct}.

In this paper, we develop the underlying formalism that proves the validity of this Dirac-bracket framework~\cite{Gonzo:2024zxo,Alessio:2025flu}.
In our formalism, the massive bodies are treated differently from the radiative degrees of freedom, which remain field-theoretic in the classical limit.
For the massive particles, we employ the Stratonovich-Weyl (SW) correspondence between the operator and phase-space formulations of quantum mechanics \cite{Weyl:1927vd,Wigner:1932eb,Groenewold:1946kp,Moyal:1949sk,stratonovich1957distributions,Gracia-Bondia:1988jyp,Varilly:1989sv,Carinena:1989uw,Sanchez:2007sn,JOLT_2018_28_4_a6}.
This correspondence is implemented by an operator function on the phase space called the SW quantizer.
It may be used to go from the Hilbert to phase space and back, although here in \sec{sec:SW} we focus on using it only in the former direction.
Note that by itself the quantizer does not apply the classical limit, albeit makes it more intuitive to do so.
In particular, point-particle states may be represented directly by Dirac delta distributions on the phase space, implying that expectation values of quantum operators are given by their SW symbols.

In \sec{sec:ClassicalObservables}, we apply the SW correspondence to scattering observables in the same way that the KMOC formalism is set up field-theoretically.
We also use the exponential representation~\cite{Bjerrum-Bohr:2021wwt,Damgaard:2021ipf} of the S-matrix $\smash{\hat{S}=e^{i\hat{K}/\hbar}}$, thereby bridging multiple approaches and deriving the classical Magnus scattering functions, from which observables may be obtained via classical Dirac brackets~\cite{Gonzo:2024zxo,Alessio:2025flu}.

Finally, in \sec{sec:MagnusAmplitudes}, we turn to the perturbative construction of Magnus amplitudes, which are matrix elements of the S-matrix logarithm~$\hat{K}$.
Recent studies have investigated their diagrammatic expansions at tree and loop level using properties of the Magnus series, revealing general organizing principles and an underlying Hopf-algebraic structure~\cite{Kim:2024svw,Brandhuber:2025igz,Kim:2025ebl,Guo:2026xaw}.
In this work, we instead derive Magnus amplitudes directly from the ordinary transition-matrix amplitudes and formulate this construction as an algorithm for computing the coefficients in their diagrammatic expansion.

\section{Stratonovich-Weyl quantization}
\label{sec:SW}

In this section, we review the SW quantization framework in quantum mechanics and extend it to the QFT setting in which additional massless particles are present.

\subsection{Stratonovich-Weyl quantizer}
\label{sec:SWquantizer}

We start by introducing the quantum-mechanical SW quantizer~$\hat\Omega(\xi)$.
Assume a classical-mechanical phase space~${\cal P}$, which is symmetric under a group~$G$ of phase-space transformations~$g$ and is equipped with a $G$-invariant phase-space measure $\dd\mu$.
Then the quantizer is defined~\cite{Carinena:1989uw} as a operator-valued function on~${\cal P}$ the obeying the following properties:
\begin{subequations}
\begin{align}
\label{eq:HermiticityAxiom}
\hat\Omega^\dagger(\xi) & = \hat\Omega(\xi) & \text{(hermiticity),} \\
\label{eq:UnitTraceAxiom}
\Tr\hat\Omega(\xi) & = 1 \hfill & \text{(unit-trace normalization),} \\
\label{eq:CovarianceAxiom}
\hat{U}(g) \hat\Omega(\xi) \hat{U}^\dagger(g) & = \hat\Omega(g(\xi)) & \text{(covariance),} \\
\label{eq:TracialityAxiom}
\smash{\int_{\cal P}}\dd\mu(\eta) \Tr[\hat\Omega(\xi) \hat\Omega(\eta)] \hat\Omega(\eta) & = \hat\Omega(\xi) & \text{(traciality).}
\end{align}
The final crucial axiom is the \emph{Stratonovich-Weyl correspondence}, which posits that the map from the operator space to that of functions on ${\cal P}$, called Weyl symbols,
\begin{equation}
\hat{A} ~\mapsto~ A_\W(\xi) := \Tr[\hat{A} \hat\Omega(\xi)]
\label{eq:WeylCorrespondenceAxiom}
\end{equation}
\label{eq:QuantizerAxioms}%
\end{subequations}
should be bijective.
These axioms imply the inverse correspondence
\begin{equation}
A_\W ~\mapsto~ \hat{A} = \int_{\cal P}\!\dd\mu(\xi) A_\W(\xi) \hat\Omega(\xi) .
\label{eq:InverseCorrespondenceProperty}
\end{equation}
It may be proven by checking that the images of the left- and right-hand side under the bijective map~\eqref{eq:WeylCorrespondenceAxiom} coincide:
\begin{align}
\!\big[\!\int\!\!\dd\mu(\xi) A_\W(\xi) \hat\Omega(\xi)\big]_\W(\eta)
:=\!\int\!\!\dd\mu(\xi) A_\W(\xi) \Tr[\hat\Omega(\xi) \hat\Omega(\eta)] &
 =\!\int\!\!\dd\mu(\xi) \Tr[\hat{A} \hat\Omega(\xi)] \Tr[\hat\Omega(\xi) \hat\Omega(\eta)] \nn \\
 = \Tr\!\big[\hat{A}\!\int\!\!\dd\mu(\xi) \hat\Omega(\xi) \Tr[\hat\Omega(\xi) \hat\Omega(\eta)]\big] &
 = \Tr[\hat{A} \hat\Omega(\eta)] =: A_\W(\eta) .
\label{eq:InverseCorrespondenceProof}
\end{align}
Therefore, the quantizer provides a clear connection between the formulation of quantum mechanics based on operators on Hilbert space ${\cal H}$ and that based on quasi-distribution functions on the phase space.
Here are some other useful properties of this correspondence:%
\begin{subequations} \begin{align}
\label{eq:ConjugationProperty}
(A^\dagger)_\W & = A_\W^*(\xi) , & \text{(conjugation),} \\
\label{eq:NormalizationProperty}
(\identity)_\W & = 1 \qquad \Leftrightarrow \qquad \int_{\cal P}\!\dd\mu(\xi) \hat\Omega(\xi) =  \identity & \text{(unity \& normalization),} \\
\label{eq:CovarianceProperty}
A_\W(g(\xi)) & = [\hat{U}^\dagger(g) \hat{A}\,\hat{U}(g)]_\W(\xi) & \text{(covariance),} \\
\label{eq:BilinearFormProperty}
\Tr[\hat{A}\hat{B}] & = \int_{\cal P}\!\dd\mu(\xi) A_\W(\xi) B_\W(\xi) & \text{(bilinear form),} \\
\label{eq:DeltaFunctionProperty}
\forall\,F(\xi) & = \int_{\cal P}\!\dd\mu(\eta) \Tr[\hat\Omega(\xi)\hat\Omega(\eta)] F(\eta) & ({\cal P}\text{ delta function),}
\end{align}
\label{eq:QuantizerProperties}%
\end{subequations}
all of which are straightforward to show from the above.

In this paper, we employ the following relativistic one-particle quantizer for each compact object of mass~$m$:\footnote{For brevity, we use
the reduced notation for momentum measures and delta functions
to absorb appropriate powers of $2\pi$; in particular,
$\ddd\!^4k := \dd^4k/(2\pi)^4$ and $\del\!_+(k^2-m^2) := 2\pi\theta(k^0)\delta(k^2-m^2)$.
The Lorentz-invariant completeness relation for scalar one-particle states then takes the form
\begin{equation*}
\identity =\!\int\!\ddd^4k\,\del\!_+(k^2-m^2) \ket{k} \bra{k}
 =\!\int\!\!\frac{\dd^3k}{(2\pi)^3 2\sqrt{\vec{k}^2+m^2}} \ket{k} \bra{k} .
\label{eq:CompletenessRelation1}
\end{equation*}
\label{foot:reduced}
}
\begin{subequations}
\begin{align}
\label{eq:RelQuantizerRkk}
\hat\Omega(x,p) := &\;2^3\!\!\int\!\ddd^4k\,\del_+(k^2\!-m^2) \big[\tfrac{p \cdot k}{m^2}\big]^{3/2} e^{i(R_p k - k) \cdot x/\hbar} \ket{R_p k} \bra{k} \\
\label{eq:RelQuantizerkRk}
= &\;2^3\!\!\int\!\ddd^4k\,\del_+(k^2\!-\!m^2) \big[\tfrac{p \cdot k}{m^2}\big]^{3/2} e^{-i(R_p k - k) \cdot x/\hbar} \ket{k} \bra{R_p k} \\
\label{eq:RelQuantizerpq}
 = & \int\!\ddd^4q\,\del(2p \cdot q) \sqrt[^4]{1{-}\tfrac{q^2}{4m^2}}\,e^{-i q \cdot x/\hbar}
   \Big|\sqrt{1{-}\tfrac{q^2}{4m^2}}p-\frac{q}{2}\Big\rangle \Big\langle\sqrt{1{-}\tfrac{q^2}{4m^2}}p+\frac{q}{2}\Big| .
\end{align} \label{eq:RelQuantizer}%
\end{subequations}
Here $R_p$ is the Grossmann-Royer reflection~\cite{Grossmann:1975ux,Royer:1977zz,Varilly:1989sv} with respect to momentum $p$.
\begin{equation}
(R_p k)^\mu := \frac{2p\!\cdot\!k}{p^2} p^\mu - k^\mu ,
\label{eq:GrossmanRoyer}
\end{equation}
which is a covariantized version of 3-momentum reflection~$k \to (k^0,-\vec{k})$.
\Eqn{eq:RelQuantizerRkk} is a more explicitly Lorentz-covariant rendition of the quantizer proposed in \cite{Carinena:1989uw} (see also \cite{Sanchez:2007sn,JOLT_2018_28_4_a6}).
For completeness, we provide explicit proofs of its defining properties as a SW quantizer in \app{app:quantizer}.

\subsection{Groenewold product and Moyal bracket}
\label{sec:starproduct}

The inverse correspondence~\eqref{eq:InverseCorrespondenceProperty} allows us to translate operator multiplication into an associative product of Weyl symbols.
For two operators $\hat{A}$ and $\hat{B}$, their Groenewold product~\cite{Groenewold:1946kp} is defined by
\begin{equation}
(A_\W \star B_\W)(\xi)
:= (\hat{A}\hat{B})_\W(\xi)
 = \smash{\int_{\cal P}}\dd\mu(\eta) \smash{\int_{\cal P}}\dd\mu(\zeta)\,
   T(\xi,\eta,\zeta)\,A_\W(\eta)\,B_\W(\zeta),
\label{eq:GroenewoldProduct}
\end{equation}
where
\begin{equation}
T(\xi,\eta,\zeta) := \Tr\!\big[\hat\Omega(\xi)\hat\Omega(\eta)\hat\Omega(\zeta)\big]
\label{eq:TrikernelDefinition}
\end{equation}
is the Stratonovich-Weyl trikernel.
The antisymmetric part of the Groenewold product is the Moyal bracket~\cite{Moyal:1949sk}
\begin{equation}
\{A_\W,B_\W\}_{\rm M}
:= [\hat{A},\hat{B}]_\W
 = A_\W \star B_\W-B_\W \star A_\W.
\label{eq:MoyalBracket}
\end{equation}
The quantum commutator thus gives rise to a bracket of functions on the phase space.

\subsection{Example: nonrelativistic particle}

To better illustrate the formalism, let us apply it to the familiar case of unconstrained three-dimensional flat space with canonical coordinates $(\vec{x}, \vec{p})$.
In this case, it is well known that the quantizer is given by the Grossmann-Royer reflection operators~\cite{Grossmann:1975ux,Royer:1977zz,Gracia-Bondia:1988jyp},
\begin{equation}
\label{eq:NRquantizer}
\hat\Omega_0(\vec{x},\vec{p}) = 2^3\!\int\!\ddd^3 k e^{-2i\vec{x}\cdot(\vec{p}-\vec{k})/\hbar} \ket{2\vec{p}{-}\vec{k}} \bra{\vec{k}}
 =\!\int\!\ddd^3 q\,e^{i\vec{q}\cdot\vec{x}/\hbar} \Big|\vec{p}-\frac{\vec{q}}{2}\Big\rangle \Big\langle\vec{p}+\frac{\vec{q}}{2}\Big| .
\end{equation}
For this example, we use the nonrelativistic momentum states normalized according to $\braket{\vec{p}\:\!|\vec{q}\:\!} = \del^{(3)}(\vec{p}-\vec{k})$.
One can then show that \eqn{eq:NRquantizer} satisfies the SW axioms~\eqref{eq:QuantizerAxioms}, where the covariance is understood to be with respect to the Galilean group.
The SW transform of an operator $\hat{A}$ reduces to the original Wigner transform~\cite{Wigner:1932eb}
\begin{equation}
A_\W(\vec{x},\vec{p})
= \Tr\! \big[\hat{A}\,\hat\Omega_0(\vec{x},\vec{p})\big]
=\!\int\!\ddd^3 q\,e^{i\vec{q}\cdot\vec{x}/\hbar} \Big\langle\vec{p}+\frac{\vec{q}}{2}\Big| \hat{A} \Big|\vec{p}-\frac{\vec{q}}{2}\Big\rangle .
\label{eq:NRWeylTransform}
\end{equation}
In our conventions, the phase-space measure is $\dd\mu(x,p) = {\dd^3 x\,\dd^3 p}/{(2\pi\hbar)^3}$, so the inverse transform is given by
\begin{equation}
\hat{A} =\!\int\!\frac{\dd^3 x\,\dd^3 p}{(2\pi\hbar)^3}\,A_\W(\vec{x},\vec{p})\ \hat\Omega_0(\vec{x},\vec{p}) .
\label{eq:NRWeylTransformInv}
\end{equation}
One can recognize that this is the usual Weyl quantization \cite{Weyl:1927vd}.

According to \eqref{eq:TrikernelDefinition}, it is straightforward to compute the explicit form of the trikernel,
\begin{equation}
\label{eq:NRtrikernel}
T_0(\vec{x},\vec{p};\vec{y}_1,\vec{k}_1;\vec{y}_2,\vec{k}_2)
 = 2^6 \exp\!\bigg(\frac{2i}{\hbar} \big[
   \vec{x} \cdot (\vec{k}_1 - \vec{k}_2)
   + \vec{y}_1 \cdot (\vec{k}_2 - \vec{p})
   + \vec{y}_2 \cdot (\vec{p} - \vec{k}_1) \big]\!\bigg) ,
\end{equation}
which gives the Groenewold product in the form of an integral transform~\eqref{eq:GroenewoldProduct}.
In this special case, it can also be represented in the more familiar differential formulation
\begin{equation}
A_\W \star B_\W
 = A_\W \exp\!\bigg(\frac{i \hbar}{2} \Big[ \cev{\partial}_{x} \cdot \vec{\partial}_{p} - \cev{\partial}_{p} \cdot \vec{\partial}_{x} \Big]\!\bigg)
B_\W ,
\label{eq:FlatGroenewoldProduct}
\end{equation}
where the arrows indicate the direction of differentiation.
The Moyal bracket is then given by the antisymmetrization of this product.
Assuming that in the small-$\hbar$ limit $A_\W$ and $B_\W$ are finite, the Moyal bracket gives the canonical Poisson bracket as the leading term,
\begin{equation}
\label{eq:Moyal2Poisson}
\{A_\W, B_\W\}_{\rm M}\, \underset{\hbar \to 0}{\longrightarrow}\, i \hbar \{ A_\W, B_\W \}_{\rm P}
\end{equation}
as expected from the SW correspondence principle.

\subsection{Relativistic point particle}

Let us return to the relativistic one-particle quantizer~\eqref{eq:RelQuantizer}, which gives the following SW transform for an operator $\hat{A}$
\begin{equation}
\label{eq:WeylTransform}
A_\W(x,p) =\!\int\!\ddd^4q\,\del(2p \cdot q) \sqrt[^4]{1{-}\tfrac{q^2}{4m^2}}\,e^{-iq \cdot x/\hbar}
   \Big\langle\sqrt{1{-}\tfrac{q^2}{4m^2}}p+\frac{q}{2}\Big|  \hat{A} \Big|\sqrt{1{-}\tfrac{q^2}{4m^2}}p-\frac{q}{2}\Big\rangle ,
\end{equation}
Here the momentum~$p$ is understood to be on shell, $p^2=m^2$, and the prefactors of \scalebox{0.75}{$\sqrt{1-\tfrac{q^2}{4m^2}}$} inside the momentum states also ensure the onshellness of the corresponding momenta.
The multiplicative factor \scalebox{0.75}{$\sqrt[4]{1-\frac{q^2}{4m^2}}$} is necessary to satisfy the traciality axiom~\eqref{eq:TracialityAxiom}, as discussed in \app{app:quantizer} during the derivation of the related delta-function property~\eqref{eq:DeltaFunctionProperty}.
Furthermore, it is important to note that the quantizer~\eqref{eq:RelQuantizerpq} enjoys a gauge shift symmetry
\begin{equation}
\label{eq:ShiftSym}
\hat\Omega(x,p) = \hat\Omega(x+\tau p,p) \qquad \Rightarrow \qquad
A_\W(x, p) = A_\W(x + \theta p, p) ,
\end{equation}
where $\theta$ could be an arbitrary function of $(x,p)$.
This means that the points on the same straight line $x= x_0 + \theta p$ are identical from the point of view of the quantizer.
In other words, the quantizer is defined as a function on the equivalence class of points $(x,p)\sim(x+\theta p,p)$.
In this sense, the relativistic quantizer is tailored to free particles.
This works well for obtaining asymptotic scattering observables, since asymptotic scattering states can be considered as free.
From a worldline perspective, the shift symmetry can be understood as the freedom to choose the origin of the worldline parameter $\tau$ along the particle trajectory $x = x_0 + \tau p/m$, which is a residual symmetry of the reparameterization gauge symmetry of the worldline action.

This shift symmetry is also crucial for the consistency of the formalism.
The physical phase space of a relativistic particle is six-dimensional, while the covariant phase space parametrized by $(x,p)$ is naively eight-dimensional.
One redundant degree of freedom is removed by the mass-shell constraint $p^2=m^2$, while the other is taken care of by the shift symmetry.
Strictly speaking, the physical one-particle phase space is given by
\begin{equation}
{\cal P} =\{(x,p)\,|\,p^2=m^2,\ p^0>0\}/\{(x,p)\sim(x+\theta p,p),\,\theta\in\mathbb{R}\} .
\end{equation}
Therefore, the covariant phase-space measure can be schematically written as
\begin{equation}
\dd\mu(x,p) = \frac{\dd^4 x\,\dd^4 p}{(2\pi\hbar)^3} \frac{\delta_+(p^2 - m^2)}{{\rm Vol}(G_\theta)} ,
\end{equation}
where ${\rm Vol}(G_\theta)$ denotes the volume of the one-parameter shift symmetry group~\eqref{eq:ShiftSym}, which is infinite.
This parallels the quotient of ${\rm PSL}(2,\mathbb{C})$ conformal Killing group in string-theoretic amplitude calculations~\cite{Tong:2009np}.
In practical calculations, we always choose a gauge-fixing condition $g(x,p)=0$, and then use the Faddeev-Popov method to obtain a finite dimensionless gauge-fixed measure
\begin{equation}
\label{eq:GaugeFixedMeasure}
\dd\mu(x,p) = 2 \frac{\dd^4 x\,\dd^4 p}{(2\pi\hbar)^3} \bigg| p \cdot \frac{\partial g(x,p)}{\partial x} \bigg| \delta\big(g(x,p)\big) \delta_+(p^2 - m^2) .
\end{equation}
This works for an arbitrary gauge constraint.
For example, we can choose $ g(x,p) = x^0$, which corresponds to the gauge choice of the worldline parameter $\tau$ as the coordinate time~$x^0$ and breaks covariance.
In this case, the measure reduces to
\begin{equation}
\label{eq:gauge1}
g(x,p) = x^0 \qquad \Rightarrow \qquad
\dd\mu(x,p) = \frac{\dd^4 x\,\dd^4 p}{(2\pi \hbar)^3} \delta(x^0) \delta\big(p^0-\sqrt{\vec{p}\,^2+m^2}\big) .
\end{equation}
Upon integrating out the time components, we recover precisely the usual nonrelativistic measure seen \eqn{eq:NRWeylTransformInv},
since the prefactor $2$ in the gauge-fixed measure~\eqref{eq:GaugeFixedMeasure} was chosen to cancel the factor of $1/2$ coming from the Lorentz-invariant on-shell delta function.

An alternative convenient choice for the gauge constraint is $g(x,p) = x \cdot p$. In this case, the measure becomes
\begin{equation}
\label{eq:gauge2}
g(x,p) = x \cdot p \qquad \Rightarrow \qquad
\dd\mu(x,p) = \frac{\dd^4 x\,\dd^4 p}{(2\pi \hbar)^3} 2m^2 \delta(x \cdot p) \delta_+(p^2 - m^2) .
\end{equation}
For two-body scattering, a more familiar gauge choice
\begin{equation}
\label{eq:gauge3}
g_1(x_1,p_1,x_2,p_2) = (x_2 - x_1) \cdot p_1, \qquad
g_2(x_1,p_1,x_2,p_2) = (x_2 - x_1) \cdot p_2 ,
\end{equation}
gives
\begin{equation}
\dd\mu = 4m_1^2 m_2^2 (\gamma^2-1)\!\prod_{j=1,2}\!\frac{\dd^4x_j \dd^4p_j}{(2\pi \hbar)^3} \delta_+(p_j^2 - m_j^2)\,\delta \big((x_2-x_1){\cdot} p_j \big) ,
\end{equation}
where $\gamma = p_1 \cdot p_2/(m_1 m_2)$ is the widely used kinematic variable (relative Lorentz factor).

With the phase-space measure defined, we can write down the Groenewold product for the relativistic quantizer. For two operators $\hat{A}$ and $\hat{B}$, we have
\begin{equation}
A_\W \star B_\W (x,p) =\!\int\!\bigg( \prod_{j=1,2}\!\dd\mu(x_j,p_j) \bigg)
T(x,p;x_1,p_1;x_2,p_2)\,A_\W (x_1,p_1)\,B_\W(x_2,p_2) .
\label{eq:R0star}
\end{equation}
As explained before, the covariant phase space has a nontrivial geometry and gauge symmetry, so the Groenewold product does not admit a simple differential representation. We present the explicit form of $T(x,p;x_1,p_1;x_2,p_2)$ in \eqn{eq:TrikernelAppendix}.

Hereafter, we will focus on relativistic point particles.
For simplicity, we consider one-particle phase space and stress the generalization to two-body phase space when necessary.

\subsection{Dirac bracket}
\label{sec:bracket}

We now turn to the classical limit of the Moyal bracket.
By the correspondence principle, this is the regime in which the angular-momentum scale set by the impact parameter and momentum is large compared with $\hbar$, schematically $|p|\cdot|b|\gg\hbar$.
As $\hbar\to0$, the Moyal bracket reduces to a classical bracket.
In an unconstrained phase space, this would be the canonical Poisson bracket, as we have seen in \eqn{eq:Moyal2Poisson}, whereas in the present relativistic setting it becomes the Dirac bracket on the physical worldline phase space.
The purpose of this section is to show how this bracket emerges directly from the small-$\hbar$ expansion of the Groenewold product.

We start from the relativistic Groenewold product~\eqref{eq:R0star} induced by the quantizer~\eqref{eq:RelQuantizer}.
Its exact integral representation in terms of the trikernel is not especially illuminating for our present purpose, but it makes the classical limit transparent.
In the small-$\hbar$ limit, the integral is controlled by the diagonal saddle $(y_1,k_1)=(x,p)$ and $(y_2,k_2)=(x,p)$ of the oscillatory trikernel.
On a fixed gauge slice, the corresponding stationary-phase expansion takes the explicit form
\begin{equation}
A_\W \star B_\W
 = A_\W B_\W
 + \frac{i\hbar}{2}\{A_\W,B_\W\}_{\rm D}
 + {\cal O}(\hbar^2) ,
\label{eq:StarDiracExpansion}
\end{equation}
where the bracket on the right-hand side, extracted from the first nontrivial term in the $\hbar$ expansion, is exactly the Dirac bracket in that gauge.
Thus the Dirac bracket is already encoded in the quantum phase-space product.
In particular, in the two-body scattering gauge~\eqref{eq:gauge3} it is precisely the Dirac bracket, whose explicit component form may be found in~\cite{Gonzo:2024zxo,Alessio:2025flu}.
A self-contained derivation from the gauge-fixed trikernel representation is relegated to \app{app:trikernel}.
Antisymmetrizing, we obtain the corresponding Moyal bracket,
\begin{equation}
\{A_\W,B_\W\}_{\rm M}
 = A_\W \star B_\W-B_\W \star A_\W
 = i\hbar\,\{A_\W^{\rm cl},B_\W^{\rm cl}\}_{\rm D} + {\cal O}(\hbar^2),
\label{eq:Moyal2Dirac}
\end{equation}
where $A_\W^{\rm cl}$ and $B_\W^{\rm cl}$ denote the leading classical parts of the corresponding Weyl symbols.

\subsection{Partial Stratonovich-Weyl transform with radiation}
\label{sec:RadiativeEffects}

For massive point particles coupled to gravity or electromagnetism, scattering observables may typically also involve massless radiation.
So far the SW transform for a relativistic massive particle has acted only on the massive Hilbert space.
With the radiative degrees of freedom included, the full Hilbert space becomes
\begin{equation}
{\cal H}={\cal H}_{\rm mass}\otimes{\cal H}_{\rm rad} .
\end{equation}
The classical limit of the massless sector is inherently field-theoretic, so we choose not to apply the SW transform to the radiative modes and restrict it to the massive degrees of freedom.
For an operator $\hat{A}$ on ${\cal H}_{\rm mass}\otimes{\cal H}_{\rm rad}$, we define the partial SW transform
\begin{equation}
\hat{A}_\W(b,p)
:= \Tr\nolimits_{{\cal H}_{\rm mass}}\!\big[ \big( \hat\Omega(b,p) \otimes \identity_{\rm rad} \big) \hat{A} \big] ,
\label{eq:PartialWeylTransform}
\end{equation}
where $\hat\Omega(b,p)$ acts only on the massive Hilbert space.
Therefore, the Weyl symbol $\hat{A}_\W$ is now a function on the massive phase space that is valued in the space of operators on ${\cal H}_{\rm rad}$.

To be more explicit, let us define the connected matrix elements of $\hat{A}_\W$ as\footnote{The prefactor $\hbar^{(m+n)/2}$ is introduced in \eqn{eq:RadiativeWeylCoefficients} so that the SW transform of the hermitian scattering operator $\hat{K}_\W^{(m,n)}$ defined below in \eqn{eq:ClassMagnusAmplitudes} is finite in the classical limit.
}
\begin{align}
\label{eq:RadiativeWeylCoefficients}
& A_\W^{(m,n)}(b,p;k_1,\ldots,k_m;k'_1,\ldots,k'_n)
:= \hbar^{(m+n)/2} \braket{k_1,\ldots,k_m|\hat{A}_\W(b,p)|k'_1,\ldots,k'_n}_{\rm c} \\ &\!
 = \hbar^{\frac{m+n}{2}}\!\!\int\!\!\ddd^4q\,\del(2p\!\cdot\!q)
   \sqrt[4]{1{-}\tfrac{q^2}{4m^2}}\,e^{-\frac{i}{\hbar} q \cdot b}
   \Big\langle\!\scalebox{0.75}{$\sqrt{1{-}\frac{q^2}{4m^2}}$}p+\frac{q}{2},k_1,\ldots,k_m\Big| \hat{A}
   \Big| \scalebox{0.75}{$\sqrt{1{-}\frac{q^2}{4m^2}}$}p-\frac{q}{2},k'_1,\ldots,k'_n\!\Big\rangle_{\!\rm c} . \nn
\end{align}
Here the subscript ``${\rm c}$'' stands for ``the connected part of'', which removes free propagation of the external massless states.
For simplicity, we suppress the discrete quantum numbers of the massless bosons.
In terms of the above matrix elements, we can write an expansion of the Weyl-symbol operator $\hat{A}_\W(b,p)$ in the massless creation and annihilation operators:
\begin{equation} \begin{aligned}
\hat{A}_\W(b,p) = \sum_{m,n=0}^\infty \frac{1}{\hbar^{(m+n)/2}m!n!} \int\!
   \Big(\prod_{i=1}^m \ddd^4k_i\,\del\!_+(k_i^2)\,a_{k_i}^\dagger\Big)
   \Big(\prod_{j=1}^n \ddd^4k_j'\,\del\!_+(k_j'^2)\,a_{k'_j}\Big) & \\ \times
   A_\W^{(m,n)}(b,p;k_1,\ldots,k_m;k'_1,\ldots,k'_n) & ,
\label{eq:RadiativeWeylExpansion}
\end{aligned} \end{equation}
The sector $(0,0)$ corresponds to the ordinary massive SW transform, namely
\begin{equation}
A_\W^{(0,0)}(b,p) = \prescript{}{\rm rad}{\braket{0|\hat{A}_\W(b,p)|0}}_{\rm rad}
 = \Tr\nolimits_{{\cal H}_{\rm mass}}\!\big[ \hat\Omega(b,p) \prescript{}{{\rm rad}}{\braket{0|\hat{A}|0}}_{\rm rad} \big] ,
\end{equation}
where $\ket{0}_{\rm rad} \in {\cal H}_{\rm rad}$ denotes the radiative vacuum.

The product of two operators maps to the Groenewold product in the massive variables, together with ordinary operator multiplication in the radiative Fock space:
\begin{equation}
(\hat{A}\hat{B})_\W = \hat{A}_\W \star \hat{B}_\W.
\end{equation}
Consequently, the transformed commutator is the mixed Moyal/commutator bracket
\begin{equation} [\hat{A},\hat{B}]_\W = \hat{A}_\W\star\hat{B}_\W - \hat{B}_\W\star\hat{A}_\W .
\label{eq:MixedMoyalCommutator}
\end{equation}
For c-number symbols, this reduces to the usual Moyal bracket on the massive phase space.
For the radiative creation and annihilation operators, it acts as the ordinary quantum commutator.

\section{Classical observables}
\label{sec:ClassicalObservables}

In this section, we proceed to derive classical scattering observables from those in QFT.

\subsection{Expectation value of point-particle state}
\label{sec:ExpectationValues}

We prepare an asymptotic initial state described by the density matrix $\hat{\rho}\in {\cal H}_{\rm mass}$.
Since we are interested in scattering of classical point particles, the relevant states should become localized on the phase space in the classical limit.
We implement this limit semiclassically, by taking $\hbar \to 0$ at fixed classical kinematics.

It is convenient to formulate this localization condition in the Wigner representation.
Let $\rho_\W$ denote the SW transform of $\hat{\rho}$, \ie its Wigner function~\cite{Wigner:1932eb}.
Within this subsection, let us consider a generic phase space $\cal P$.
For simplicity, first consider a one-particle incoming state localized near a phase-space point $\xi$ in the classical limit.
The point-particle approximation then means that $\rho_{\W,\xi}$ becomes sharply supported at~$\xi$: for any sufficiently regular phase-space function $F$,
\begin{equation}
\lim_{\hbar \to 0} \int_{\cal P}\!\dd\mu(\eta)\rho_{\W,\xi}(\eta) F(\eta) = F(\xi) .
\end{equation}
This implies that the classical limit of the Wigner function is given by the delta function in the distributional sense
\begin{equation}
\rho_{\W,\xi}(\eta) \xrightarrow[\hbar\to0]{} \delta_{\cal P}(\eta,\xi) .
\label{eq:ClassicalRho}
\end{equation}
Let us consider an operator $\hat{O}$.
Its expectation value for the state $\hat{\rho}$ is
\begin{equation}
\braket{\hat{O}}
 = \langle 0 | \Tr\nolimits_{{\cal H}_{\rm mass}}\!\big[ \hat{O} \hat{\rho} \big] | 0 \rangle
 = \int_{\cal P}\!\dd\mu(\eta) \langle 0 | \hat{O}_\W(\eta) \, \rho_{\W,\xi}(\eta) | 0 \rangle ,
\end{equation}
where $\ket{0}$ denotes the vacuum of the radiative sector.
Provided that the SW transform of $\hat{O}$ has a smooth classical limit, using \eqn{eq:ClassicalRho}, we see that the expectation value in this localized state reduces in the classical limit to its Weyl symbol evaluated at~$\xi$:
\begin{equation}
\int_{\cal P}\!\dd\mu(\eta) \hat{O}_\W(\eta)\,\rho_{\W,\xi}(\eta)
~\xrightarrow[\hbar\to0]{}~ \hat{O}_\W(\xi)
\qquad \Rightarrow \qquad
\braket{\hat{O}}_{\rm cl} = \lim_{\hbar \to 0} O^{(0,0)}_\W(\xi) .
\end{equation}

Specializing to the relativistic one-particle case, let the initial state be represented by the covariant pair $(b,p)$, or more precisely by its equivalence class on the physical phase space.
This corresponds to an asymptotic free trajectory $x = b + p\tau/m$, where $b$ is the impact parameter relative to the origin.
Inserting the explicit quantizer~\eqref{eq:RelQuantizer} gives
\begin{equation}
O^{(0,0)}_\W(b,p) = \int\!\ddd^4q\,\del(2p \cdot q)
   \sqrt[^4]{1{-}\tfrac{q^2}{4m^2}}\,e^{-i q \cdot b/\hbar}
   \Big\langle\sqrt{1{-}\tfrac{q^2}{4m^2}}p+\frac{q}{2}\Big| \hat{O}
   \Big|\sqrt{1{-}\tfrac{q^2}{4m^2}}p-\frac{q}{2}\Big\rangle .
\end{equation}
We note that $O^{(0,0)}_\W(b,p)$ still depends on $\hbar$, and the above expression should be understood as evaluating $\hat{O}$ on a point-particle-like state.

For two-body scattering, the initial state is represented by $(b_1,p_1;b_2,p_2)$ on the two-body phase space, so that $b:=b_1-b_2$ is the impact parameter.
According to the correspondence principle, the classical scattering regime is the large-angular-momentum limit, \ie
$|\vec{b}\times\vec{p}_{1,2}|\gg\hbar$ in the center-of-mass frame.

\subsection{Hermitian scattering operator}
\label{sec:Noperator}

Classical scattering observables are usually defined as differences between outgoing and incoming operators, such as momentum impulses or spin kicks \cite{Kosower:2018adc,Maybee:2019jus,delaCruz:2020bbn,Cristofoli:2021vyo,Aoude:2021oqj,Britto:2021pud,Cristofoli:2021jas}.
Let $\hat{O}$ denote an appropriate observable of the incoming state, and let $\hat{S}$ denote the scattering operator.
The change of the observable under scattering is given by the Heisenberg-picture transformation
\begin{equation}
\label{eq:ObservableChange}
\Delta \hat{O}:=\hat{S}^\dagger \hat{O}\hat{S}-\hat{O}.
\end{equation}
Usually, we will expand the S-matrix as  $\hat{S} = \identity + i \hat{T}/\hbar$, where $\hat{T}$ is the transition operator that gives the usual scattering amplitude.
However, it is well known that for a $2 \to 2$ process, the scattering amplitude contains superclassical terms in the classical limit~\cite{Kosower:2018adc}.
Alternatively, it has been noted~\cite{Bjerrum-Bohr:2021wwt,Damgaard:2021ipf} that the logarithm of the S-matrix is convenient for extracting the classical limit.
We therefore write the scattering operator in the form
\begin{equation}
\hat{S}=\exp\!\left(\frac{i}{\hbar}\hat{K}\right),
\label{eq:HermitianScatteringOperator}
\end{equation}
Unitarity of $\hat{S}$ implies that $\hat{K}$ is hermitian,
\begin{equation}
\hat{K}^\dagger=\hat{K},
\end{equation}
whose connected matrix elements are often called \emph{Magnus amplitudes}.
For more details on properties of Magnus amplitudes, see~\cite{Bjerrum-Bohr:2021wwt,Damgaard:2021ipf, Damgaard:2023ttc,Gonzo:2024zxo, Kim:2024svw, Alessio:2025flu, Kim:2025gis, Kim:2025hpn, Kim:2025olv, Haddad:2025cmw, Kim:2025sey, Kim:2025ebl, Brandhuber:2025igz, Gonzo:2026yha, Guo:2026xaw}.
These amplitudes are expressed in terms of diagrams with retarded, advanced, and Hadamard cut propagators, weighted by factors that are known as Murua coefficients~\cite{Kim:2024svw,Kim:2025ebl,Brandhuber:2025igz,Guo:2026xaw}.
In \sec{sec:MagnusAmplitudes}, we provide a straightforward algorithm to compute these diagram coefficients based on expanding the Magnus amplitudes in terms of scattering amplitudes.

Plugging \eqn{eq:HermitianScatteringOperator} into \eqn{eq:ObservableChange} and using the Baker-Campbell-Hausdorff formula then gives the operator identity
\begin{equation}
\Delta \hat{O}
= \sum_{r=1}^\infty \frac{(-i)^r}{\hbar^r r!}
\underbrace{[\hat{K},[\hat{K},\ldots,[\hat{K},\hat{O}]]]}_{r\text{-fold commutator}}.
\label{eq:ObservableBCH}
\end{equation}

Let us now consider the scattering of two massive point particles, parametrized by the initial phase-space point $(b_1,p_1;b_2,p_2)$.
It is then easy to see that taking the partial SW transform over the massive-particle Hilbert spaces and using \eqn{eq:MixedMoyalCommutator} yields the observable operator on the radiative Fock space
\begin{equation}
\Delta \hat{O}_\W
=
\sum_{r=1}^\infty \frac{(-i)^r}{\hbar^r r!}
\underbrace{\Big\{\hat{K}_\W,\big\{\hat{K}_\W,\ldots\{\hat{K}_\W,\hat{O}_\W\}\!\ldots\big\}\Big\}}_{r\text{-fold mixed Moyal/commutator bracket}} ,
\label{eq:ObservableOperator}
\end{equation}
where the braces denote the mixed bracket of \eqref{eq:MixedMoyalCommutator}.
Restricting to two-body scattering without incoming radiation means taking the vacuum expectation value of the above operator-valued Weyl symbol.
To extract its classical limit, we need to understand the behavior of the partially transformed Magnus amplitudes $K_\W^{(m,n)}$ in the small-$\hbar$ limit.
That said, the same equation~\eqref{eq:ObservableOperator} may be used for processes with incoming radiation, such as Compton scattering~\cite{Akhtar:2025nmt,Ivanov:2026icp,Brunello:2026lzf} (in which case the SW transform is for a single massive body).

\subsection{Classical observables from Magnus amplitudes}
\label{sec:observables}

To derive classical observables, the classical limit of the Moyal bracket must be supplemented by that of the partially transformed connected matrix elements $K_\W^{(m,n)}$ of the hermitian scattering operator.
This requires the standard multi-soft scaling, in which all internal and external massless momenta are taken to scale as $\hbar$ \cite{Kosower:2018adc}.
The behavior of the corresponding $\hat{K}$-matrix elements in this limit was analyzed in~\cite{Bjerrum-Bohr:2021wwt}.
In our conventions, $K_\W^{(m,n)}$ is normalized so that this limit is finite:
\begin{equation} \begin{aligned}
 & \chi^{(m,n)}(b_1,p_1;b_2,p_2;\bar{k}_1,\ldots,\bar{k}_m;\bar{k}'_1,\ldots,\bar{k}'_n) := \\ & \quad
   \lim_{\hbar\to 0}
   K_\W^{(m,n)}(b_1,p_1;b_2,p_2;\hbar\bar{k}_1,\ldots,\hbar\bar{k}_m;\hbar\bar{k}'_1,\ldots,\hbar\bar{k}'_n) ,
\label{eq:ClassMagnusAmplitudes}
\end{aligned} \end{equation}
where $\bar{k}_j^\mu:=k_j^\mu/\hbar$ are wavevectors, \ie rescaled massless momenta.
For simplicity, we also write $\chi := \chi^{(0,0)}$ for the particular case of the radial action~\cite{Damgaard:2023ttc}.
More generally, we tend to refer to these objects as ``classical Magnus amplitudes'' despite them not being entirely in momentum space.
Correspondingly, we define the classical operator-valued function $\hat{\chi}$ by replacing $K_\W^{(m,n)}$ with $\chi^{(m,n)}$ in $\hat{K}_\W$,
\begin{equation}
\label{eq:ClassicalOperator}
\hat{\chi}(b_1,p_1;b_2,p_2) = \sum_{m,n=0}^{\infty} \frac{1}{m! n!}
    \int \prod_{i=1}^{m} \Big( \ddd^4\bar{k}_i\,\del\!_+(\bar{k}_i^2)\,\bar{a}_{\bar{k}_i}^\dagger \Big)
    \prod_{j=1}^{n} \Big( \ddd^4\bar{k}_j'\,\del\!_+(\bar{k}_j'^2)\,\bar{a}_{\bar{k}_j^\prime} \Big) \chi^{(m,n)} ,
\end{equation}
where the $\hbar$ scaling has been absorbed into
\begin{equation}
\bar{a}_{\bar{k}} = \hbar^{\frac32} a_{\hbar \bar{k}}, \qquad \quad
\bar{a}_{\bar{k}}^\dagger = \hbar^{\frac32} a_{\hbar \bar{k}}^\dagger .
\end{equation}
We further define the classical bracket of the creation and annihilation operators by
\begin{equation}
\{\bar{a}_{\bar{k}}, \bar{a}_{\bar{k}^\prime}^\dagger\} := -\frac{i}{\hbar} [\bar{a}_{\bar{k}}, \bar{a}_{\bar{k}^\prime}^\dagger] = -2 i \omega_{\bar{k}} \del^{(3)}(\bar{k} - \bar{k}^\prime).
\end{equation}
With these definitions, the classical limit of two-body observables can be written in the compact form
\begin{equation}
\Delta O
= \lim_{\hbar \to 0} \Delta O_\W^{(0,0)}(b_1,p_1;b_2,p_2)
= \sum_{r=1}^{\infty}\frac{1}{r!}
\big\langle 0\big|
\underbrace{\Big\{ \hat{\chi}, \big\{ \hat{\chi}, \big\{ \hat{\chi}, \ldots \{ \hat{\chi}, \hat{O}_\W \}\!\ldots\big\}\big\}\Big\}}_{r\text{-fold mixed Dirac/commutator bracket}}
\big|0\big\rangle ,
\label{eq:RadiativeObservableMaster}
\end{equation}
with $|0\rangle$ the vacuum of the massless Hilbert space.
By \eqn{eq:Moyal2Dirac}, whenever the quantity entering the mixed bracket is a c-number function on the reduced massive phase space, its classical limit is governed by the Dirac bracket derived in \sec{sec:bracket}.

\subsection{Global observables}
\label{sec:globalobs}

We first consider observables that do not act on the radiative Fock space,
\begin{equation}
\hat{O}_\W=O(b_1,p_1;b_2,p_2)\,\identity .
\end{equation}
This includes, in particular, the momentum impulse obtained by choosing $O=p_1^\mu$ or~$p_2^\mu$.
Since vacuum expectation values annihilate any unmatched creation or annihilation operator, the leading contribution to \eqn{eq:RadiativeObservableMaster} is controlled by the radial-action sector $\chi^{(0,0)}=:\chi$.
We immediately find
\begin{equation}
\Delta O^{(1)}
=
\big\{\chi,O\big\}_{\rm D}.
\label{eq:GlobalObsLeading}
\end{equation}
Restricting to the nonradiative sector, the conservative contribution is given by nested Dirac brackets generated by $\chi$,
\begin{equation}
\Delta O\big|_{\rm con} = \sum_{k=1}^\infty \frac{1}{k!} \underbrace{\Big\{ \chi, \big\{ \chi, \ldots \{ \chi, O
\}\!\ldots\big\}\Big\}}_{k\text{-fold Dirac bracket}}(b_1,p_1;b_2,p_2) .
\end{equation}
The first radiative effect appears at the subleading order.
Evaluating the double mixed bracket and projecting onto the (radiative) vacuum yields
\begin{equation}\!\!
\Delta O^{(2)}\big|_{\rm rad} =
-\frac{i}{2}\int\!\ddd^4\bar{k}\,\del\!_+(\bar{k}^2)\,
\Big(
\chi^{(0,1)}(\bar{k})\,\big\{\chi^{(1,0)}(\bar{k}),O\big\}_{\rm D}
-\chi^{(1,0)}(\bar{k})\,\big\{\chi^{(0,1)}(\bar{k}),O\big\}_{\rm D}
\Big) .
\label{eq:GlobalObsQuadratic}
\end{equation}
This arises from the contraction of a single-emission Magnus amplitude with a single-absorption one.
Higher orders are obtained in the same way and organize themselves into nested Dirac brackets involving the radiative classical Magnus amplitudes~$\chi^{(m,n)}$ with more massless legs.

\subsection{Waveform}
\label{sec:waveformsubsec}

The waveform is obtained by taking the observable itself to be a radiative mode operator. In the far zone, with $r:=|\vec{x}|$, retarded time $u:=t-r$, and $q^\mu=\hbar\omega(1,\hat{n})$, the metric perturbation is~\cite{Alessio:2025flu}
\begin{equation}
\langle \kappa h_{\mu\nu}(x) \rangle
=
\frac{\kappa}{4\pi r}
\sum_{\sigma}\int\!\frac{d\omega}{2\pi}
\Big(
e^{-i\omega u}\varepsilon^\sigma_{\mu\nu}(\hat{n})\,\langle \bar{a}_{\bar{q},\sigma}\rangle
+e^{i\omega u}\varepsilon^{*\sigma}_{\mu\nu}(\hat{n})\,\langle \bar{a}_{\bar{q},\sigma}^\dagger\rangle
\Big) .
\label{eq:WaveformFarZone}
\end{equation}
It is therefore enough to evaluate $\langle \bar{a}_{\bar{q},\sigma} \rangle$, with the conjugate term following by hermiticity.
We omit the polarization label from here on for brevity.

At leading order, only the single-emission Magnus amplitude contributes, and \eqn{eq:RadiativeObservableMaster} gives
\begin{equation}
\langle \bar{a}_{\bar{q}} \rangle^{(1)}
= \langle 0|\{ \hat{\chi}, \bar{a}_{\bar{q}} \} |0\rangle
= i \chi^{(1,0)}(\bar{q}) .
\label{eq:aLeading}
\end{equation}
The next term already exhibits the interplay between conservative evolution and genuinely radiative channels:
\begin{equation}
\langle \bar{a}_{\bar{q}} \rangle^{(2)}
=
\frac{i}{2}
\big\{\chi,\chi^{(1,0)}(\bar{q})\big\}_{\rm D}
+\frac{1}{2}\!\int\!\ddd^4\bar{k}\,\del\!_+(\bar{k}^2)
\Big[
\chi^{(0,1)}(\bar{k})\,\chi^{(2,0)}(\bar{q},\bar{k})
-\chi^{(1,0)}(\bar{k})\,\chi^{(1,1)}(\bar{q};\bar{k})
\Big] .
\label{eq:aQuadratic}
\end{equation}
We see that the second term captures interference with intermediate radiative states, similarly to the cut contributions in the KMOC formalism~\cite{Kosower:2018adc}.
Substituting \eqns{eq:aLeading}{eq:aQuadratic} into \eqn{eq:WaveformFarZone} produces the waveform order by order directly from the partially transformed hermitian operator, confirming the conjecture in~\cite{Alessio:2025flu}.

\section{Magnus amplitudes from transition matrix}
\label{sec:MagnusAmplitudes}

In perturbation theory, Magnus amplitudes can be computed using diagrams similar to ordinary scattering-amplitude Feynman diagrams.
Their propagators, however, are not simply Feynman propagators.
We therefore refer to these diagrams as \emph{Magnus diagrams}.

\subsection{Murua coefficients from transition matrix}
\label{sec:MuruaCoefficients}

At tree level, Kim et al.~\cite {Kim:2024svw} show that each internal propagator of a Magnus diagram can be assigned a retarded or advanced prescription, and the resulting terms are weighted by graph functions known as Murua coefficients~\cite{murua2006hopf}. 
At loop level, Brandhuber et al.~\cite{Brandhuber:2025igz} find that the propagators could also be Hadamard cut functions.
A diagrammatic algorithm for computing the corresponding Murua coefficients is given in~\cite{Guo:2026xaw}, where the authors also expand Magnus diagrams in a color basis~\cite{Kim:2025ebl}.

These constructions follow from the Magnus series expansion of the hermitian operator~$\hat{K}$~\cite{magnus1954exponential}. 
An alternative approach, which we adopt here, is to determine Magnus amplitudes directly from their relation to ordinary transition amplitudes~\cite{Damgaard:2021ipf,Damgaard:2023ttc}. 
This provides an algorithm for expressing Magnus diagrams in terms of Feynman and oriented-cut propagators, which are Wightman functions. 
These propagators can subsequently be rewritten in terms of retarded, advanced, and symmetric cut propagators, namely Hadamard functions. 
The coefficients multiplying the resulting terms are precisely the Murua coefficients.

We begin by expressing $\hat{K}$ in terms of the transition operator $\hat{T}$:
\begin{equation}
\label{eq:KinT}
    \frac{i}{\hbar} \hat{K}
    = \log \bigg(\identity + \frac{i}{\hbar} \hat{T} \bigg)
    = \sum_{j=1}^{\infty}\frac{(-1)^{j+1}}{j} \bigg(\frac{i \hat{T}}{\hbar} \bigg)^j.
\end{equation}
When evaluating a matrix element of $\hat{K}$, we insert completeness relations between successive transition operators,
\begin{equation}
    \bigg(\frac{i \hat{T}}{\hbar}\bigg)^j =
    \frac{i \hat{T}}{\hbar}
    \Big(\sum_{n}\big| n \big\rangle \big\langle n \big| \Big)
    \frac{i \hat{T}}{\hbar}
    \Big(\sum_{n}\big| n \big\rangle \big\langle n \big| \Big) \dots
    \Big(\sum_{n}\big| n \big\rangle \big\langle n \big| \Big)
    \frac{i \hat{T}}{\hbar}.
\end{equation}
Here we use a schematic completeness relation
\begin{equation}
    \identity = \sum_{n}\big| n \big\rangle \big\langle n \big|.
\end{equation}
For example, the completeness relation for scalar one-particle states is given in \foot{foot:reduced}.
Feynman and oriented cut propagators that arise from these intermediate-state insertions are
\begin{equation}
\Delta_{\rm F}(p) = \frac{i}{p^2 - m^2 + i \varepsilon}, \qquad \quad
\Delta_{\rm c}(p) = \del_+(p^2 - m^2).
\end{equation}

Let us consider a Magnus diagram with $v$ vertices.
Each vertex simply comes from the usual Feynman rules, but may originate from one of the multiple transition operators $\hat{T}$ in \eqn{eq:KinT}.
We should include all possible ways to assign the vertices to the $\hat{T}$ matrices.
Propagators connecting vertices belonging to the same $\hat{T}$ matrix element are Feynman propagators. 
Propagators connecting vertices belonging to different transition operators are oriented cut propagators. 
Their direction is determined by the ordering of the corresponding $\hat{T}$ operators.

Since every factor of $\hat{T}$ must contain at least one vertex, the sum in \eqn{eq:KinT} truncates at $j=v$. 
More explicitly, for each $j=1,\ldots,v$, we enumerate all ordered partitions of the set of vertices into $j$ nonempty subsets. 
For a given ordered partition, propagators within the same subset are assigned the Feynman prescription, while propagators between different subsets are assigned oriented-cut prescriptions. 
Summing over all ordered partitions and multiplying the contribution with $j$ subsets by $(-1)^{j+1}/j$ gives the complete Magnus contribution.

The Feynman and oriented-cut propagators can be rewritten in terms of retarded and symmetric cut propagators (also known as the Hadamard propagator):
\begin{equation}
\begin{aligned}
\Delta_{\rm F}(p) & = \frac{\Delta_{\rm r}(p) + \Delta_{\rm r}(-p)}{2} + \Delta_{\rm H}(p) , \\
\Delta_{\rm c}(p) & = \frac{\Delta_{\rm r}(p) - \Delta_{\rm r}(-p)}{2} + \Delta_{\rm H}(p) , 
\end{aligned}  \qquad \text{with} \qquad
\begin{aligned}
&\Delta_{\rm r}(p) := \frac{i}{p^2 - m^2 + i\varepsilon p^0} , \\
&\Delta_{\rm H}(p) := \frac{1}{2} \del(p^2-m^2).
\end{aligned}
\label{eq:Propagators}
\end{equation}
After substituting these relations and collecting terms with identical propagator assignments, one obtains the retarded/Hadamard representation of the Magnus diagram.
The coefficients of the individual terms are exactly the Murua coefficients.

\subsection{Sample Murua-coefficient computation}
\label{sec:SampleMuruaCoefficients}

As a simple illustration, consider a diagram with three vertices
\begin{equation}
\raisebox{-0.55\height}{\includegraphics{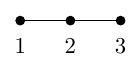}}
\label{eq:SampleDiagram}
\end{equation}
with only the propagators shown explicitly.
Note that the external legs may connect to any vertex but are omitted.
The ordered partitions of the above diagram are
\begin{subequations} \begin{align}
    j=1:& \quad \{1,2,3\}, \\
    j=2:& \quad
        \{1\}{\leftarrow} \{2,3\},~
        \{2\}{\leftarrow} \{1,3\},~
        \{3\}{\leftarrow} \{1,2\}, \\ &\quad
        \{2,3\}{\leftarrow} \{1\},~
        \{1,3\}{\leftarrow} \{2\},~
        \{1,2\}{\leftarrow} \{3\}, \nn \\
    j=3:& \quad
       \{1\}{\leftarrow} \{2\}{\leftarrow} \{3\},~
       \{2\}{\leftarrow} \{3\}{\leftarrow} \{1\},~
       \{3\}{\leftarrow} \{1\}{\leftarrow} \{2\}, \\ &\quad
       \{1\}{\leftarrow} \{3\}{\leftarrow} \{2\},~
       \{2\}{\leftarrow} \{1\}{\leftarrow} \{3\},~
       \{3\}{\leftarrow} \{2\}{\leftarrow} \{1\}. \nn
\end{align} \label{eq:SamplePartitions}
\end{subequations}
For example, the ordered partition
$\{1\}{\leftarrow}\{2,3\}$ gives
\begin{equation}
\Delta_{\rm c}(p_{12}) \Delta_{\rm F}(p_{23}) = \raisebox{-0.55\height}{\includegraphics{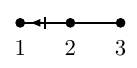}} ,
\end{equation}
where the arrow on a cut propagator indicates its direction and the momenta are labeled by the vertices that they connect.
For instance, $\Delta_{\rm c}(p_{ab})$ is the oriented cut propagator with the momentum~$p_{ab}$ flowing from vertex $b$ to vertex $a$.

Summing over all contributions from the ordered partitions~\eqref{eq:SamplePartitions} of the above diagram and plugging in \eqn{eq:Propagators} gives
\begin{align}\!\!\!
\raisebox{-0.6\height}{\includegraphics[scale=0.95]{diagram/three_points_plain.pdf}}\!
    &= \frac{1}{3} \Delta_{\rm r}(p_{12}) \Delta_{\rm r}(p_{23})
    +\frac{1}{3} \Delta_{\rm r}(p_{21}) \Delta_{\rm r}(p_{32})
    +\frac{1}{6} \Delta_{\rm r}(p_{12}) \Delta_{\rm r}(p_{32})
    +\frac{1}{6} \Delta_{\rm r}(p_{21}) \Delta_{\rm r}(p_{23}) \nn \\[-5mm]
    &=\frac{1}{3} \raisebox{-0.6\height}{\includegraphics[scale=0.95]{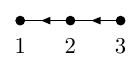}}
    +\,\frac{1}{3} \raisebox{-0.6\height}{\includegraphics[scale=0.95]{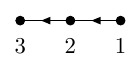}}
    +\,\frac{1}{6} \raisebox{-0.42\height}{\includegraphics[scale=0.95]{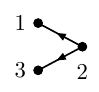}}
    +\,\frac{1}{6} \raisebox{-0.42\height}{\includegraphics[scale=0.95]{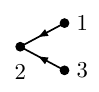}} .
\label{eq:diagram_example}
\end{align}
The arrows indicate the causal flow associated with the retarded propagators.

We have not included any diagrammatic symmetry factors in this example; they can be restored in the usual way.
For instance, if vertices $1$ and $3$ are attached to identical external legs, the diagram carries a symmetry factor of $1/2$, and the first two terms in \eqn{eq:diagram_example} are the same.
The corresponding contribution to $i\hat{K}/\hbar$ is then
\begin{equation}
\frac{1}{2} \raisebox{-0.6\height}{\includegraphics{diagram/three_points_plain.pdf}}\!
    =\frac{1}{3} \raisebox{-0.6\height}{\includegraphics{diagram/three_points_retarded_2to1_3to2.pdf}}
    +\,\frac{1}{12} \raisebox{-0.42\height}{\includegraphics{diagram/three_points_retarded_2to1_2to3.pdf}}
    +\,\frac{1}{12} \raisebox{-0.42\height}{\includegraphics{diagram/three_points_retarded_1to2_3to2.pdf}} .
\end{equation}

\subsection{Implementation and benchmark}

The above presentation may merely look like a detailed definition of the Murua coefficients in the context of Magnus amplitude calculations.
However, it also constitutes a viable computational algorithm whose direct
implementation is already competitive with other approaches.
It therefore provides a direct and systematic procedure for extracting Murua coefficients from ordinary transition-amplitude diagrams.

We provide a straightforward {\tt Mathematica} implementation in an ancillary file.
The computed coefficients agree with the available ancillary data of~\cite{Guo:2026xaw} (tagged as ``BW'' for the ``black-and-white'' basis) --- after accounting for the minor discrepancy in the propagator conventions~\eqref{eq:Propagators}.\footnote{Our Murua coefficients $\omega(G)$ are related to those of Guo et al.~\cite{Guo:2026xaw}, denoted $\omega_{\rm Guo}(G)$ via
\begin{equation*}
\omega_{\rm Guo}(G)=(-1)^{E+L/2} \omega(G),
\end{equation*}
where $E$ and $L$ are the numbers of propagators and loops, respectively.
}
The performance of our code is compared in \tab{tab_muruaret_wbwms_benchmark} with that of the implementation supplied with~\cite{Guo:2026xaw}.

We note that the timing comparison requires some care, since the two algorithms have different native outputs.
First, Guo et al.~\cite{Guo:2026xaw} suppress the cut propagators, so the benchmark is restricted to the sector with no cut propagators.
Second, our algorithm generates all propagator assignments in a single run, including both retarded orientations and the Hadamard-cut option for each edge.
By contrast, the implementation of~\cite{Guo:2026xaw} computes one graph at a time.
We therefore compare our one-run timing with the cumulative time for all non-isomorphic representatives.\footnote{The implementation of~\cite{Guo:2026xaw} caches the coefficients of intermediate graphs.
For each reported timing, the cache is initially empty and shared across the $N_{\rm iso}$ evaluations.
}
This also suggests a possible optimization of our present implementation:
the ordered-partition sum could be projected onto a prescribed propagator assignment, thereby avoiding the full expansion when only a single graph is required.

These representative timings were obtained on a 2023 Macbook Pro with Apple M2 Pro processor (2.42--3.5 GHz, 12 cores) and 16~GB RAM, with no attempt at code optimization on our part. 
The timings in \tab{tab_muruaret_wbwms_benchmark} should therefore be regarded as a baseline for a deliberately
transparent implementation rather than an optimized limit.

\begin{table}[t]
  \centering
  \small
  \begin{tabular}{@{}rrrccccc@{}}
    \toprule
    & & & \multicolumn{2}{c}{This work}
        & \multicolumn{2}{c}{Guo et al.~\cite{Guo:2026xaw}} & Ratio \\
    \cmidrule(lr){4-5}\cmidrule(lr){6-7}
    $V$ & $L$ & $E$
      & $2^E$ & Time [s]
      & $N_{\rm iso}$ & Time [s]
      & $T_{\rm Guo}/T$ \\
    \midrule
    4 & 0 & 3 &   8 & 0.00343 &   4 &  0.00470 &  1.37  \\
    5 & 0 & 4 &  16 & 0.0212  &  10 &  0.0194  &  0.912 \\
    6 & 0 & 5 &  32 & 0.160   &  16 &  0.0489  &  0.305 \\
    7 & 0 & 6 &  64 & 1.68    &  36 &  0.249   &  0.148 \\
    4 & 2 & 5 &  32 & 0.00675 &   6 &  0.0472  &  7.00  \\
    5 & 2 & 6 &  64 & 0.0526  &  21 &  0.449   &  8.52  \\
    6 & 2 & 7 & 128 & 0.486   &  45 &  3.63    &  7.47  \\
    7 & 2 & 8 & 256 & 4.18    & 105 & 23.3     &  5.56  \\
    5 & 4 & 8 & 256 & 0.226   &  18 &  0.943   &  4.18  \\
    6 & 4 & 9 & 512 & 2.59    & 180 & 59.8     & 23.0  \\
    \bottomrule
  \end{tabular}
  \caption{Representative timings in the retarded/advanced propagator sector. 
  Here $(V,L,E)$ denote the numbers of vertices, loops and propagators, respectively. 
  For each row, we select the first Magnus diagram in the corresponding BW ancillary data file of~\cite{Guo:2026xaw}.
  Our time is for a single run producing all $2^E$ labeled orientations of propagators (suppressing the cut propagators), whereas the time for~\cite{Guo:2026xaw} is the cumulative time for evaluating all $N_{\rm iso}$ non-isomorphic representatives graph by graph.}
  \label{tab_muruaret_wbwms_benchmark}
\end{table}

\section{Conclusion}
\label{sec:outro}

In this paper, we have formulated a partially phase-space framework that connects scattering data in QFT to the classical dynamics of relativistic massive bodies, while retaining massless radiation as a field-theoretic sector (as opposed to introducing a functional-derivative bracket~\cite{Peierls:1952cb}).
The central ingredient is a covariant Stratonovich-Weyl quantizer~\eqref{eq:RelQuantizer} for relativistic one-particle states.
We have shown that classical expectation values of field-theoretic observable operators reduce directly to their Weyl symbols evaluated on the corresponding asymptotic trajectory.

In the classical limit, we consider the logarithm of the $\hat{S}$ matrix, $\hat{K} := -i\hbar\log \hat{S}$, whose partial SW transform gives rise to the eikonal-like classical Magnus amplitudes~$\chi^{(m,n)}$ that serve as generating functions of classical dynamics with $m+n$ external radiative legs, as defined in \eqn{eq:ClassMagnusAmplitudes}.
The nested-bracket structure induced by quantum commutation relations provides a neat way to extract classical observables from $\chi^{(m,n)}$.
This approach meshes well between the KMOC formalism and eikonal methods in the spirit of \cite{Damgaard:2023ttc} but with the advantage of the integro-differential Poisson/Dirac bracket machinery (as opposed to a purely integral one).

Finally, we describe a method for obtaining Magnus amplitudes from ordinary transition matrix elements.
We implement this algorithm in an ancillary file and verify that it efficiently extracts the Murua coefficients, particularly for loop diagrams.

We expect the proposed formalism to have many applications and extensions, which we leave for future work.
In fact, a complementary concurrent work~\cite{Moynihan:2026tbd} proposing a similar approach focuses on multiple interesting observables relevant for gravitational-wave physics.
There are certain discrepancies in the way the relativistic Weyl symbols are introduced here and in \cite{Moynihan:2026tbd}, but many classical-physics applications may be insensitive to such details.

In particular, we expect our framework to extend naturally to spinning bodies, for which the phase-space constraints and associated geometry are more nontrivial.
In this case, it should provide a rigorous justification for the Dirac-bracket approach to the spinning two-body problem~\cite{Gonzo:2024zxo,Alessio:2025flu,Kim:2024svw,Kim:2024grz,Kim:2025hpn,Kim:2025olv}. 
This approach is well suited for describing classical gravitational dynamics with arbitrary spin orientations and has already found exciting black-hole scattering applications~\cite{Akpinar:2025bkt,Akpinar:2025tct}.

\begin{acknowledgments}
We thank Francesco Alessio and Riccardo Gonzo for enlightening conversations and collaboration on related projects,
as well as Nathan Moynihan for letting us know about his concurrent work~\cite{Moynihan:2026tbd} following CS's talk at Amplitudes 2026.
We are grateful to Junran Zheng for useful discussions about the algorithm for calculating Murua coefficients.
We also thank Sixun Du and Jung-Wook Kim for helpful conversations.
AO is supported by the Science and Technology Commission of the Shanghai Municipality via STCSM grant No. 24ZR1450600.
CS is supported by the China Postdoctoral Science Foundation under Grant No. 2022TQ0346 and the National Natural Science Foundation of China under Grant No. 12347146.
\end{acknowledgments}

\appendix
\section{Relativistic quantizer properties}
\label{app:quantizer}

In this appendix, we verify that the properties~\eqref{eq:QuantizerAxioms} hold for the operator~\eqref{eq:RelQuantizer}, which establishes it as the relativistic SW quantizer.

First of all, let us point out the equivalence of the two representations~\eqref{eq:RelQuantizerRkk} and~\eqref{eq:RelQuantizerpq}, which relies on the integration-variable change
\begin{equation}
q = k-R_p k \qquad \Leftrightarrow \qquad k = \sqrt{1{-}\tfrac{q^2}{4m^2}}\,p+\frac{q}{2} ; \qquad
p \cdot k = m^2 \sqrt{1{-}\tfrac{q^2}{4m^2}} ,
\label{eq:k2q}
\end{equation}
with the Jacobian converting neatly between the formulas:
\begin{equation}
\det\!\bigg(\!2\frac{\partial\vec{k}}{\partial\vec{q}}\bigg)
 = 1 + \frac{q^0}{2p^0\sqrt{1{-}\tfrac{q^2}{4m^2}}}
\quad \Rightarrow \quad
2^3\ddd^4k\,\del_+(k^2\!-\!m^2) \big[\tfrac{p \cdot k}{m^2}\big]^{3/2}\!
 = \ddd^4q\,\del(2p \cdot q) \sqrt[^4]{1{-}\tfrac{q^2}{4m^2}} .
\label{eq:k2qJacobian}
\end{equation}
The rewriting~\eqref{eq:RelQuantizerpq} is nice because it makes the hermiticity property~\eqref{eq:HermiticityAxiom} of the quantizer manifest
(namely conjugation corresponds to $q\to -q$ integration-variable change).
Of course, hermiticity implies the third equivalent representation~\eqref{eq:RelQuantizerkRk}.

The variant~\eqref{eq:RelQuantizerpq} is also convenient for proving the trace property~\eqref{eq:UnitTraceAxiom}.
One may simply use the inner product
\begin{equation}
\Big\langle\sqrt{1{-}\tfrac{q^2}{4m^2}}p+\frac{q}{2}\Big|\sqrt{1{-}\tfrac{q^2}{4m^2}}p-\frac{q}{2}\Big\rangle
  = 2\bigg(\!\sqrt{1{-}\tfrac{q^2}{4m^2}}p^0+\frac{q^0}{2}\bigg) \del^{(3)}(\vec{q}\:\!) ,
\end{equation}
which sets $\vec{q}$ (and therefore also $q^0$) to zero and gives
\begin{equation}
\Tr\hat\Omega(x,p)
 =\!\int\!\dd q^0\,\delta(2p^0 q^0) \sqrt[^4]{1{-}\tfrac{(q^0)^2}{4m^2}}\,e^{-i q^0 x^0/\hbar}\,
   2\bigg(\!\sqrt{1{-}\tfrac{(q^0)^2}{4m^2}}p^0+\frac{q^0}{2}\bigg) = 1 .
\end{equation}

Next, we choose to prove the following version of the delta-function property~\eqref{eq:DeltaFunctionProperty}:
\begin{equation}
\Tr[\hat\Omega(x,p)\hat\Omega(y,k)]
 = \hbar^3\,\del^{(3)}(\vec{p}-\vec{k}\:\!)\,\delta^{(3)}\big(\vec{x}-\vec{y}-[x^0\!-y^0]\vec{p}/p^0\big) .
\label{eq:DeltaFunctionPropertyRel}
\end{equation}
Using the same representation~\eqref{eq:RelQuantizerpq} for both quantizers, we rewrite the left-hand side as
\begin{equation} \begin{aligned}
\int\!\frac{\ddd^3 q\,\ddd^3 q'}{4p^0k^0} \sqrt[^4]{1{-}\tfrac{q^2}{4m^2}} \sqrt[^4]{1{-}\tfrac{q'^2}{4m^2}}\,e^{-i(q \cdot x + q'\cdot y)/\hbar}
   \Big\langle\sqrt{1{-}\tfrac{q^2}{4m^2}}p+\frac{q}{2}\Big|\sqrt{1{-}\tfrac{q'^2}{4m^2}}k-\frac{q'}{2}\Big\rangle & \\ \times
   \Big\langle\sqrt{1{-}\tfrac{q'^2}{4m^2}}k+\frac{q'}{2}\Big|\sqrt{1{-}\tfrac{q^2}{4m^2}}p-\frac{q}{2}\Big\rangle &,
\end{aligned} \end{equation}
where leave $\smash{q^0 = \vec{p}\cdot\vec{q}/p^0}$, $\smash{q'^0 = \vec{k}\cdot\vec{q}\;\!'\!/k^0}$, $\smash{p^0=\sqrt{\vec{p}\;\!^2+m^2}}$ and $\smash{k^0=\sqrt{\vec{k}^2+m^2}}$ implicit.
The two inner products above are equal to
\begin{equation} \begin{aligned}
 4 \bigg(\!\sqrt{1{-}\tfrac{q^2}{4m^2}}p^0+\frac{q^0}{2}\bigg)
   \bigg(\!\sqrt{1{-}\tfrac{q^2}{4m^2}}p^0-\frac{q^0}{2}\bigg)
   \del^{(3)}&\bigg(\!\sqrt{1{-}\tfrac{q^2}{4m^2}}\vec{p}+\frac{\vec{q}}{2}
                    - \sqrt{1{-}\tfrac{q'^2}{4m^2}}\vec{k}+\frac{\vec{q}\:\!'}{2} \bigg) \\ \times
   \del^{(3)}&\bigg(\!\sqrt{1{-}\tfrac{q'^2}{4m^2}}\vec{k}+\frac{\vec{q}\;\!'}{2}
                    - \sqrt{1{-}\tfrac{q^2}{4m^2}}\vec{p}+\frac{\vec{q}}{2} \bigg) .
\label{eq:DeltaFunctionPropertyInnerProducts}
\end{aligned} \end{equation}
Changing the second integration variable to $\vec{Q}:=\vec{q}+\vec{q}\;\!'$, we find the following nontrivial delta-function identity
\begin{equation} \begin{aligned}
\del^{(3)}\!\bigg( \frac{\vec{Q}}{2} + \sqrt{1{-}\tfrac{q^2}{4m^2}}\vec{p} - \sqrt{1{-}\tfrac{q'^2}{4m^2}}\vec{k} \bigg)
\del^{(3)}\!\bigg( \frac{\vec{Q}}{2} + \sqrt{1{-}\tfrac{q'^2}{4m^2}}\vec{k} - \sqrt{1{-}\tfrac{q^2}{4m^2}}\vec{p} \bigg) & \\
 = \Big(1{-}\tfrac{q^2}{4m^2}\Big)^{-1/2} \Big(1{-}\tfrac{q^2}{4m^2}{-}\tfrac{(q^0)^2}{4(p^0)^2}\Big)^{-1}
   \del^{(3)}(\vec{Q}) \del^{(3)}(\vec{p}-\vec{k}) & ,
\label{eq:DeltaFunctionPropertyInnerProducts2}
\end{aligned} \end{equation}
which leads us to
\begin{equation} \begin{aligned}
\del^{(3)}(\vec{p}-\vec{k})\!\int\!\frac{\ddd^3 q\,\ddd^3 Q}{(p^0)^2}
   \sqrt[^4]{1{-}\tfrac{q^2}{4m^2}} \sqrt[^4]{1{-}\tfrac{q'^2}{4m^2}}\,e^{-i(q \cdot x + q'\cdot y)/\hbar}
   \bigg[\Big(1{-}\tfrac{q^2}{4m^2}\Big)(p^0)^2-\frac{(q^0)^2}{4}\bigg] & \\ \times
   \Big(1{-}\tfrac{q^2}{4m^2}\Big)^{-1/2} \Big(1{-}\tfrac{q^2}{4m^2}{-}\tfrac{(q^0)^2}{4(p^0)^2}\Big)^{-1} \del^{(3)}(\vec{Q}) & .
\label{eq:DeltaFunctionPropertyCancel}
\end{aligned} \end{equation}
Here $\vec{Q}=0$ automatically sets $\vec{q}\;\!'=-\vec{q}$ and hence $q'^0=-\vec{k}\cdot\vec{q}/k^0=-\vec{p}\cdot\vec{q}/p^0=-q^0$.
In this way, we arrive at the following integral representation of the spatial delta function:
\begin{equation}
\Tr[\hat\Omega(x,p)\hat\Omega(y,k)]
 =\!\del^{(3)}(\vec{p}-\vec{k})\!\int\!\ddd^3q\,
   e^{-i\left( [x^0\!-y^0]\vec{p}\cdot\vec{q}/p^0  - \vec{q} \cdot [\vec{x}-\vec{y}] \right)/\hbar} ,
\end{equation}
which completes the proof of \eqn{eq:DeltaFunctionPropertyRel}.
Note that in order for the $\vec{q}$-dependent prefactors from \eqn{eq:DeltaFunctionPropertyInnerProducts2} to entirely cancel in \eqn{eq:DeltaFunctionPropertyCancel}, we needed the subtle integrand factor~\scalebox{0.75}{$\sqrt[4]{1-\frac{q^2}{4m^2}}$} inside the relativistic quantizer~\eqref{eq:RelQuantizerpq}.

The delta-function property~\eqref{eq:DeltaFunctionPropertyRel} implies the traciality axiom~\eqref{eq:TracialityAxiom}.
Indeed, using the generically gauge-fixed measure~\eqref{eq:GaugeFixedMeasure} while assuming $p^2=m^2$ (as usual), we obtain
\begin{align}
\label{eq:TracialityAxiomCancel}
 2\!\int&\!\frac{\dd^4 y\,\dd^4 k}{(2\pi\hbar)^3}
   \bigg| k \cdot \frac{\partial g(y,k)}{\partial y} \bigg| \delta\big(g(y,k)\big) \delta_+(k^2 - m^2)
   \Tr[\hat\Omega(x,p) \hat\Omega(y,k)] \hat\Omega(y,k) \\ &
 = \frac{1}{p^0}\!\int\!\dd^4 y \bigg| p \cdot \frac{\partial g(y,p)}{\partial y} \bigg| \delta\big(g(y,p)\big)
   \delta^{(3)}\big(\vec{x}-\vec{y}-[x^0\!-y^0]\vec{p}/p^0\big)\,\hat\Omega(y,p) \nn \\ &
 =\!\int\!\dd g(y,p)\,\dd^3\big(\vec{y}-\vec{x}-[y^0\!-x^0]\vec{p}/p^0\big)\,
   \delta\big(g(y,p)\big) \delta^{(3)}\big(\vec{y}-\vec{x}-[y^0\!-x^0]\vec{p}/p^0\big)\,\hat\Omega(y,p) , \nn
\end{align}
where in the last step we have used the following Jacobian determinant
\begin{equation}
\det\!\left(\begin{smallmatrix}
\partial g/\partial y^0 & \partial g/\partial y^1 & \partial g/\partial y^2 & \partial g/\partial y^3 \\
-p^1/p^0 & 1 & 0 & 0 \\ -p^2/p^0 & 0 & 1 & 0 \\ -p^3/p^0 & 0 & 0 & 1
\end{smallmatrix}\right) = \frac{p^\mu}{p^0} \frac{\partial g(y,p)}{\partial y^\mu} .
\end{equation}
Having thus trivialized the integration over the delta functions, we see from \eqn{eq:TracialityAxiomCancel}:
\begin{equation}\!\!\!
\int\!\!\dd\mu(y,k) \Tr[\hat\Omega(x,p) \hat\Omega(y,k)] \hat\Omega(y,k)
 = \hat\Omega\big(y^0,\vec{x}{+}[y^0\!{-}x^0]\vec{p}/p^0,p\big) \big|_{g\left(y^0\!,\vec{x}+[y^0-x^0]\vec{p}/p^0\!,p\right)=0} .\!
\end{equation}
This coincides with $\hat\Omega(x,p)$ on the support of the gauge constraint $g(x,p)=0$, since the latter should be designed specifically to fix the redundancy~\eqref{eq:ShiftSym} of the form $x^\mu \to x^\mu + \theta p^\mu$, where now we have $\theta=[y^0\!-x^0]/p^0$.
So we have verified the traciality property~\eqref{eq:TracialityAxiom} of the relativistic quantizer~\eqref{eq:RelQuantizerpq}.

Let us show how the covariance axiom~\eqref{eq:CovarianceAxiom} is implemented for the Poincar\'e transformations $x^\mu \to L^\mu{}_\nu x^\nu+a^\mu$, which are represented on the one-particle Hilbert space by the unitary operators satisfying
\begin{equation}
\hat{U}(L,a) \ket{p} = e^{ia \cdot (Lp)/\hbar} \ket{Lp} .
\end{equation}
This time using for concreteness the formula~\eqref{eq:RelQuantizerRkk}, we compute $\hat{U}(L,a) \hat\Omega(x,p) \hat{U}^\dagger(L,a)$ as
\begin{align}
\label{eq:CovarianceAxiomDerivation}
 & 2^3\!\!\int\!\ddd^4k\,\del_+(k^2\!-m^2) \big[\tfrac{p \cdot k}{m^2}\big]^{3/2}
   e^{i(R_p k - k) \cdot x/\hbar} e^{ia \cdot (L(R_p k))/\hbar} \ket{L(R_p k)}
   \bra{Lk} e^{-ia \cdot (Lk)/\hbar} \\ &
 = 2^3\!\!\int\!\ddd^4k'\,\del_+(k'^2\!-m^2) \big[\tfrac{(Lp)\cdot k'}{m^2}\big]^{3/2}
   e^{i(R_{Lp} k' - k') \cdot (Lx)/\hbar} e^{ia \cdot(R_{Lp} k')/\hbar} e^{-ia \cdot k'/\hbar}
   \ket{R_{Lp} k'} \bra{k'} , \nn
\end{align}
where we have switched to $k'^\mu:=L^\mu{}_\nu k^\nu$ and used the simple property
\begin{equation}
L^\mu{}_\nu (R_p k)^\nu = \frac{2p\!\cdot\!k}{p^2} L^\mu{}_\nu p^\nu - L^\mu{}_\nu k^\nu
 = [R_{Lp}(Lk)]^\mu
\end{equation}
of the Grossmann-Royer reflection.
Combining the exponentials of \eqn{eq:CovarianceAxiomDerivation} into the single
$\smash{e^{i(R_{Lp} k' - k') \cdot (Lx+a)/\hbar}}$, we arrive at the same quantizer representation~\eqref{eq:RelQuantizerRkk} but with transformed arguments:
\begin{equation}
\hat{U}(L,a) \hat\Omega(x,p) \hat{U}^\dagger(L,a) = \hat\Omega(Lx+a,Lp) .
\label{eq:CovarianceAxiomRel}
\end{equation}

Finally, let us note that, unlike the first four axioms~\eqref{eq:HermiticityAxiom}--\eqref{eq:TracialityAxiom}, the last axiom~\eqref{eq:WeylCorrespondenceAxiom} is not something that one should check using a proposed SW quantizer candidate, such as \eqn{eq:RelQuantizer}, but rather an underlying assumption about operators on the Hilbert space that is made explicit in the formal definition~\eqref{eq:QuantizerAxioms}.

\section{From trikernel to Dirac bracket}
\label{app:trikernel}

In this appendix, we derive the classical Dirac bracket directly from the relativistic Groenewold product.
Since the essential point is already visible for a single massive body, we present the derivation in the one-body covariant phase space.
The extension to the two-body problem is completely straightforward.

\subsection{Trikernel}
We begin by deriving the trikernel entering the relativistic Groenewold product~\eqref{eq:R0star}.
An equivalent expression was obtained in~\cite{Carinena:1989uw}; see eq.~(48) therein.
For the relativistic quantizer~\eqref{eq:RelQuantizer}, it is convenient to start from the cyclic representation
\begin{equation}
T(x_1,p_1;x_2,p_2;x_3,p_3)
:= \Tr\!\big[\hat\Omega(x_1,p_1)\hat\Omega(x_2,p_2)\hat\Omega(x_3,p_3)\big] .
\end{equation}
All three momenta lie on the same positive-energy mass shell, $p_i^2=m^2$ and $p_i^0>0$.
For the $i$th quantizer, let~$q_i$ denote the internal momentum difference in \eqn{eq:RelQuantizerRkk}.
Then
\begin{equation}
\label{eq:trikernel0}
    T = \int \prod_{i=1}^3 \bigg(\ddd^4 k_i\,\del_+(k_i^2 - m^2)\,
    2^3 \Big[\frac{p_i {\cdot} k_i}{m^2}\Big]^{3/2} e^{-iq_i\cdot x_i/\hbar} \bigg)
    \langle k_1 | R_{p_2} k_2 \rangle
    \langle k_2 | R_{p_3} k_3 \rangle
    \langle k_3 | R_{p_1} k_1 \rangle,
\end{equation}
where
\begin{equation}
    R_{p_i} k_i = 2 \frac{p_i {\cdot} k_i }{m^2} p_i - k_i, \quad
    q_i = k_i - R_{p_i} k_i, \quad
     \langle k_i | R_{p_j} k_j \rangle = 2 k_i^0 \del^{(3)} (\vec{k_i} -\! R_{p_j} \vec{k_j}) .
\end{equation}
We first perform the $k_2$ and $k_3$ integrations.
The last two inner products in \eqn{eq:trikernel0} localize these integrals according to
\begin{equation}
    \int\!\ddd^4 k_i \,\del_+(k_i^2 - m^2)
    \langle k_i | R_{p_j} k_j \rangle \, f(k_i, \dots)
    =
    f(k_i, \dots)\big|_{k_i \to R_{p_j} k_j},
\end{equation}
for an arbitrary function~$f(k_i)$.
Using the Grossmann-Royer reflections, all momenta can then be expressed in terms of the remaining integration variable~$k_1$:
\begin{gather}
    k_3 = R_{p_1} k_1 , \qquad
    k_2 = R_{p_3} k_3 = R_{p_3} R_{p_1} k_1,
    \\
    q_1 = k_1 {-} R_{p_1} k_1 , \qquad
    q_2 = R_{p_3} R_{p_1} k_1 {-} R_{p_2} R_{p_3} R_{p_1} k_1, \qquad
    q_3 = R_{p_1} k_1 {-} R_{p_3} R_{p_1} k_1. \nn
\end{gather}
After these integrations, the trikernel reduces to
\begin{equation}
T =2^9\!\int\!\ddd^4 k_1 \,\del_+(k_1^2 - m^2) \,
    \Big[\prod_{i=1}^3 \frac{p_i {\cdot} k_i}{m^2}\Big]^{3/2}
    e^{-i\sum_{i=1}^3 q_i\cdot x_i/\hbar}
    \langle k_1 | R_{p_2} R_{p_3} R_{p_1} k_1 \rangle ,
\end{equation}
while the remaining inner product is
\begin{equation}
    \langle k_1 | R_{p_2} R_{p_3} R_{p_1} k_1 \rangle = 2 k_1^{0} \del^{(3)}( \vec k_1 - R_{p_2} R_{p_3} R_{p_1} \vec k_1 ) .
\end{equation}
The support condition imposed by the remaining three-dimensional delta function has a unique solution on the positive-energy mass shell,
\begin{equation}
    k_{1\star}
    := \frac{1}{m^2\rho^{1/2}}
    \big[(p_2\cdot p_3)p_1 + (p_1\cdot p_3)p_2 -(p_1\cdot p_2)p_3 \big] ,
\end{equation}
where the dimensionless momentum invariant is
\begin{equation}
    \rho := \frac{(p_3\cdot p_1)^2+(p_1\cdot p_2)^2+(p_2\cdot p_3)^2}{m^4}
    -\frac{2(p_3\cdot p_1)(p_1\cdot p_2)(p_2\cdot p_3)}{m^6} .
    \label{eq:CDefinitionAppendix}
\end{equation}
The corresponding Jacobian is
\begin{equation}
    \left|
    \det\!\left(
    \frac{\partial\big(\vec k_1 - R_{p_2} R_{p_3} R_{p_1} \vec k_1\big)}
    {\partial \vec{k}_1}
    \right)_{\!k_1=k_{1\star}}
    \right|
    = 8 \rho .
\end{equation}
It follows that the delta function can be written as
\begin{equation}
    \del^{(3)}( \vec k_1 - R_{p_2} R_{p_3} R_{p_1} \vec k_1 )
    = \Theta(\rho) \frac{\del^{(3)}( \vec k_1 - \vec k_{1\star} )}{8\rho} .
\end{equation}
The factor~$\Theta(\rho)$ enforces the condition $\rho>0$ required for this solution.
Finally, integrating over~$k_1$ yields
\begin{align}
    \label{eq:TrikernelAppendix}
    T&(x,p; x_1, p_1; x_2,p_2)
    = \frac{2^6\Theta(\rho)}{m^9\rho^{13/4}}
    \big[(p\cdot p_1)( p_1\cdot p_2)(p_2\cdot p)\big]^{3/2}
    \\
    &\quad\times
    \exp\!\bigg\{\frac{2i}{m^2\hbar\rho^{1/2}}
    \Big[(p\cdot p_1) p_2\cdot(x-x_1)
    +(p_1\cdot p_2) p\cdot(x_1-x_2)
    +(p_2\cdot p) p_1\cdot(x_2-x)\Big]\bigg\} , \nn
\end{align}
where we have relabeled $(x_3,p_3)$ as $(x,p)$.
Under the same relabeling, $\rho$ is given by \eqn{eq:CDefinitionAppendix}.

\subsection{Local coordinates}
For a single worldline, the Groenewold product takes the form
\begin{equation}
\label{eq:star_product}
    (A \star B)(x,p) = \int\!\dd \mu(x_1, p_1) \dd \mu(x_2, p_2) T(x,p;x_1,p_1;x_2,p_2) A(x_1, p_1) B(x_2, p_2) .
\end{equation}
Unlike in the nonrelativistic case, no simple all-order differential representation is available.
The first two terms in the small-$\hbar$ expansion can nevertheless be extracted directly from \eqn{eq:star_product}.
In particular, the antisymmetric ${\cal O}(\hbar)$ term is governed by the Dirac bracket.

To parametrize the physical phase space, we first introduce independent local coordinates~$P^a$, with $a=1,2,3$, on the mass shell, so that $p^\mu=p^\mu(P^a)$. The corresponding tangent vectors are
\begin{equation}
    \label{eq:tangent_vectors}
    E_a^\mu := \frac{\partial p^\mu}{\partial P^a} \quad \text{with} \quad p_\mu E_a^\mu = 0.
\end{equation}
The induced metric is
\begin{equation}
    \label{eq:induced_metric}
    h_{ab} := E_a^\mu E_b^\nu \eta_{\mu\nu}.
\end{equation}
The indices of the induced metric are raised and lowered by $h_{ab}$ and its inverse $h^{ab}$.
We also have the completeness relation
\begin{equation}
    \label{eq:completeness}
    \Pi^{\mu\nu} := \eta^{\mu\nu} - \frac{p^\mu p^\nu}{m^2} = h^{ab} E_a^\mu E_b^\nu .
\end{equation}
The corresponding shift-invariant position coordinates are\footnote{The minus sign follows from our mostly minus metric convention.}
\begin{equation}
    X_a = -E_a^\mu x_\mu, \quad \text{with} \quad X_a(x+ \alpha p, p) = X_a(x, p).
\end{equation}
The pairs $(X_a,P^a)$ form canonical coordinates on the reduced phase space. The natural measure is therefore\footnote{For example, choosing $P^a=p^i$ reproduces the gauge-fixed measure~\eqref{eq:gauge1} obtained by the Faddeev-Popov procedure.}
\begin{equation}
    \label{eq:measure}
    \dd \mu(x,p) = \frac{\dd^3 P \dd^3 X}{(2\pi \hbar)^3},
    \quad \text{with} \quad
    \dd^3 P = \dd P^1 \dd P^2 \dd P^3,
    \quad
    \dd^3 X = \dd X_1 \dd X_2 \dd X_3 .
\end{equation}
The phase of the trikernel is invariant under $x_i \to x_i+\alpha_i p_i$ for arbitrary $\alpha_i\in\mathbb{R}$. Defining
\begin{align}
    \label{eq:YK}
    Y_{1,a} := E_a^\mu(P_1) (x_1 - x)_\mu, &\qquad
    K_{2}^a := E^a_\mu(P_1) \big[ (p_1 \!\cdot p_2) p^\mu - (p \!\cdot p_1) p_2^\mu \big], \nn \\
    Y_{2,a} := E_a^\mu(P_2) (x_2 - x)_\mu, &\qquad
    K_{1}^a := E^a_\mu(P_2) \big[ (p_2 \!\cdot p) p_1^\mu -(p_1 \!\cdot p_2) p^\mu \big],
\end{align}
we may write this phase as
\begin{equation}
    \label{eq:phase}
    \Phi = \frac{2 i }{m^2 \hbar \rho^{1/2}} \Big( K_{2}^a Y_{1,a} + K_{1}^a Y_{2,a} \Big).
\end{equation}
We next change the integration variables from $(X_{i,a},P_i^a)$ to $(Y_{i,a},P_i^a)$ for $i=1,2$. Since
\begin{equation}
    Y_{i,a} = -X_{i,a} - E_a^\mu(P_i) x_\mu ,
\end{equation}
the Jacobian has unit absolute value, and hence
\begin{equation}
    \dd \mu(x_i,p_i) \propto \dd^3 P_i \dd^3 X_i
    = \dd^3 P_i \dd^3 Y_i .
\end{equation}
Conversely, the covariant coordinates can be written as
\begin{equation}
    x_i^\mu = x^\mu + E_a^\mu(P_i) Y_i^a + \alpha_i p_i^\mu .
\end{equation}
Shift invariance allows us to set $\alpha_i=0$ without loss of generality.

The star product then becomes
\begin{align}
    \label{eq:star_product_XP}
    (A \star B)(x,p) ={}& \int \Big( \prod_{i=1,2} \frac{\dd^3 P_i \dd^3 Y_i}{(2\pi \hbar)^3} \Big)
    {\cal N} \exp\Big[\frac{2 i }{m^2 \hbar \rho^{1/2}} \Big( K_{2}^a Y_{1,a} + K_{1}^a Y_{2,a} \Big) \Big] \nn \\
    & \times A(x^\mu\!+ E_a^\mu(P_1) Y_1^a, p_1)\,B(x^\mu\!+ E_a^\mu(P_2) Y_2^a, p_2) ,
\end{align}
where ${\cal N} = \frac{2^6 \Theta(\rho)}{m^9 \rho^{13/4}} (p \cdot p_1)^{3/2} (p_1 \cdot p_2)^{3/2} (p_2 \cdot p)^{3/2}$ is the normalization factor.
The only nontrivial dependence on $\hbar$ is from the Fourier phase, and it can be absorbed by rescaling $Y_{i,a} = \hbar \bar{Y}_{i,a}$.
The small-$\hbar$ expansion is therefore generated by Taylor expanding $A$ and $B$:
\begin{subequations}
\begin{align}
    A(x_\mu + \hbar E^a_\mu(P_1) \bar{Y}_{1,a}, p_1) &= \sum_{n=0}^{\infty} \frac{\hbar^n}{n!} \left( \bar{Y}_{1,a} E^a_\mu(P_1) \frac{\partial}{\partial x_\mu} \right)^n A(x, p_1), \\
    B(x_\mu + \hbar E^a_\mu(P_2) \bar{Y}_{2,a}, p_2) &= \sum_{m=0}^{\infty} \frac{\hbar^m}{m!} \left( \bar{Y}_{2,a} E^a_\mu(P_2) \frac{\partial}{\partial x_\mu} \right)^m B(x, p_2) .
\end{align}   
\end{subequations}

\paragraph{Leading order.}
At leading order, $n=m=0$, and the integrand depends on $\bar{Y}_{i,a}$ only through the Fourier phase.
Each $\bar{Y}_i$ integration thus produces a three-dimensional delta function.
We first integrate over~$\bar{Y}_2$:
\begin{equation}
\label{eq:star_product_XP0_step1}
    \int\!\dd^3 \bar{Y}_2 \exp\Big[ \frac{2 i}{m^2 \rho^{1/2}} K_{1}^a \bar{Y}_{2,a} \Big]
    =
    \frac{m^6 \rho^{3/2}}{2^3} \del^{(3)}(K_1^a) .
\end{equation}
On the positive-energy mass shell, $K_1^a=0$ has the unique solution $p_1^\mu=p^\mu$.
Therefore,
\begin{equation}
    \del^{(3)}(K_1^a)
    =
    \frac{\del^{(3)}(P_1^a-P^a)}
    {\left|\det\!\left(\partial K_1^a/\partial P_1^b\right)\right|_{P_1=P}} .
\end{equation}
On this support,
\begin{equation}
\label{eq:solveP1}
    P_1^a = P^a
    \qquad \Rightarrow \qquad
    \begin{aligned} &
    \rho = 1, \quad
    {\cal N} = \frac{2^6 (p \cdot p_2)^{3}}{m^6}, \quad
    K_2^a = -m^2 E^a_\mu(P) p_2^\mu , \\ &
    \frac{\partial K_1^a}{\partial P_1^b}\bigg|_{P_1=P}
    = E^a_\mu(P_2) \big((p\cdot p_2)\delta^\mu_\nu - p^\mu p_{2\nu} \big) E^\nu_b(P) .
    \end{aligned}
\end{equation}
Substituting \eqns{eq:star_product_XP0_step1}{eq:solveP1} and then integrating over~$P_1^a$ gives
\begin{equation}
    \label{eq:solvedP1Y2}
    (A \star B)(x,p)\big|_{\hbar^0} = \int\!\frac{\dd^3 P_2 \dd^3 \bar{Y}_1}{(2\pi)^3}
    \frac{2^3 (p \cdot p_2)^{3}}{|\det(\partial K_1^a / \partial P_1^b)|_{P_1= P}}
    \exp\big[{-} 2 i \bar{Y}_{1,a} E^a_\mu(P) p_2^\mu \big]
    A(x, p) B(x, p_2) .
\end{equation}
The remaining $\bar{Y}_1$ integration similarly produces $\del^{(3)}\big(E^a_\mu(P) p_2^\mu\big)$:
\begin{equation}
    (A \star B)(x,p)\big|_{\hbar^0} = \int\!\frac{\dd^3 P_2}{(2\pi)^3}
    \frac{(p \cdot p_2)^{3}}{|\det(\partial K_1^a / \partial P_1^b)|_{P_1= P}}
    \del^{(3)}\big(E^a_\mu(P) p_2^\mu\big)
    A(x, p) B(x, p_2) .
\end{equation}
This delta function localizes the remaining momentum at $P_2^a=P^a$, for which
\begin{equation}
\label{eq:solveP2}
    P_2^a = P^a
    \quad \Rightarrow \quad
    (p\cdot p_2)^3 = m^{6}, \quad
    \left|\det\!\left(\frac{\partial K_1^a}{\partial P_1^b}\right)\right| = m^6, \quad
    \left|\det\!\left(\frac{\partial[E^a_\mu(P)p_2^\mu]}{\partial P_2^b}\right)\right| = 1 .
\end{equation}
Consequently, the leading term is the pointwise product
\begin{equation}
    (A \star B)(x,p)\big|_{\hbar^0} = A(x,p) B(x,p) .
\end{equation}

\paragraph{Subleading order.}
At the order linear in~$\hbar$, the pairs $(n,m)=(1,0)$ and $(0,1)$ contribute. We first evaluate the $(1,0)$ term; the other is obtained analogously. The relevant term in the Taylor expansion is
\begin{equation}
    \hbar \bar{Y}_{1,a} E^a_\mu(P_1) \frac{\partial}{\partial x_\mu} A(x, p_1) .
\end{equation}
The $\bar{Y}_2$ and $P_1$ integrations proceed exactly as at leading order, leading to an expression similar to \eqn{eq:solvedP1Y2}, with $A(x, p)$ replaced by the above term,
\begin{equation} \begin{aligned}
    \label{eq:solvedP1Y2_2}
    (A \star B)(x,p)\big|_{n=1,m=0}
    ={}& \hbar \int\!\frac{\dd^3 P_2 \dd^3 \bar{Y}_1}{(2\pi)^3}
    \frac{2^3 (p \cdot p_2)^{3}}
    {\left|\det\!\left(\partial K_1^a/\partial P_1^b\right)\right|_{P_1=P}} \\
    &\times \exp[-2 i \bar{Y}_{1,b} E^b_\mu(P) p_2^\mu]
    \Big(\bar{Y}_{1,a} E^a_\mu(P) \frac{\partial A(x, p) }{\partial x_\mu} \Big) B(x, p_2) .
\end{aligned} \end{equation}
It is useful to define
\begin{equation}
    \bar{K}_2^a := E^a_\mu(P) p_2^\mu .
\end{equation}
Changing the integration variable from $P_2^a$ to $\bar{K}_2^a$ gives
\begin{align}
    & \dd \bar{K}_2^a = E^a_\mu(P) E_b^{\mu}(P_2) \dd P_2^b \nn \\
    \Rightarrow \quad
    &\dd P_2^a = E_\mu^a(P_2) \Big[ \delta^\mu_\nu - \frac{p^\mu p_{2, \nu}}{p\cdot p_2} \Big] E_b^\nu(P) \dd \bar{K}_2^b
    = \frac{1}{p\cdot p_2}
    \left.\frac{\partial K_1^a}{\partial P_1^b}\right|_{P_1=P}
    \dd \bar{K}_2^b .
\end{align}
Therefore, the measure transforms as
\begin{align}
    \dd^3 P_2
    \frac{(p\cdot p_2)^3}
    {\left|\det\!\left(\partial K_1^a/\partial P_1^b\right)\right|_{P_1=P}}
    = \dd^3 \bar{K}_2 .
\end{align}
Within the Fourier integral, $\bar{Y}_{1,a}$ can be replaced by a derivative with respect to $\bar{K}_2^a$:
\begin{equation}
    \bar{Y}_{1,a} e^{-2i\bar{Y}_{1,b}\bar{K}_2^b}
    =
    \frac{i}{2}\frac{\partial}{\partial \bar{K}_2^a}
    e^{-2i\bar{Y}_{1,b}\bar{K}_2^b} .
\end{equation}
\Eqn{eq:solvedP1Y2_2} then becomes
\begin{equation}
(A \star B)(x,p)\big|_{n=1,m=0}
 = \frac{i\hbar}{2}\!\int\!\frac{\dd^3 \bar{K}_2 \dd^3 \bar{Y}_1}{(2\pi)^3}
   2^3 \frac{\partial}{\partial \bar{K}_2^a} \Big( e^{-2 i \bar{Y}_{1,b} \bar{K}_2^b} \Big)
   \Big( E^a_\mu(P) \frac{\partial A(x, p) }{\partial x_\mu} \Big) B(x, p_2(\bar{K}_2)) .
\end{equation}
Assuming that the boundary term vanishes, we integrate by parts so that the derivative acts on $B(x,p_2(\bar{K}_2))$. The remaining Fourier integrals impose $p_2=p$ and give
\begin{equation}
    (A \star B)(x,p)\big|_{n=1,m=0} = -\frac{i\hbar}{2}
    \left( E^a_\mu(P) \frac{\partial A(x, p) }{\partial x_\mu} \right)
    \left.
    \left(\frac{\partial B(x,p_2(\bar{K}_2))}{\partial \bar{K}_2^a}\right)_{\!X}
    \right|_{\bar{K}_2=0} .
\end{equation}
Here the derivative with respect to $\bar{K}_2^a$ is taken at fixed canonical position~$X_a = -E_a^\mu(P_2) x_\mu$, as inherited from the canonical measure~\eqref{eq:measure}. Using
\begin{equation}
    E^a_\mu(P) = -\frac{\partial x_\mu }{ \partial X_a}, \qquad
    \frac{\partial P_2^b}{\partial \bar{K}_2^a}\bigg|_{P_2 = P } = \delta^b_a ,
\end{equation}
we obtain
\begin{equation}
    (A \star B)(x,p)\big|_{n=1,m=0} = \frac{i\hbar}{2}
    \frac{\partial A(x, p)}{\partial X_a}
    \left.\frac{\partial B(x, p)}{\partial P^a}\right|_{X} .
\end{equation}
Adding the $(n,m)=(0,1)$ term gives the complete first-order contribution,
\begin{equation}
    (A \star B)(x,p)\big|_{\hbar^1} = \frac{i\hbar}{2}
    \left( \frac{\partial A(x, p)}{\partial X_a}
    \left.\frac{\partial B(x, p)}{\partial P^a}\right|_{X}
    - \left.\frac{\partial A(x, p)}{\partial P^a}\right|_{X}
    \frac{\partial B(x, p)}{\partial X_a} \right) .
\end{equation}
Antisymmetrizing the Groenewold product, we find that the leading term of the Moyal bracket is the Poisson bracket in the canonical coordinates $(X_a,P^a)$:
\begin{equation}
    \{A, B\}_{\rm M}(x,p)
    = i \hbar \{A(X_a, P^a), B(X_a, P^a)\}_{\rm P}
    + {\cal O}(\hbar^2) .
\end{equation}

\subsection{Dirac bracket}
The reduced Poisson bracket admits a covariant representation as a Dirac bracket once a gauge has been chosen. We show this in the gauge $x\cdot p=0$, together with the mass-shell constraint $p^2=m^2$. On this gauge slice,
\begin{equation}
    x_\mu = - E^a_\mu(P) X_a . 
\end{equation}
The derivatives at fixed~$X_a$ satisfy
\begin{equation}
    \frac{\partial A}{\partial X_a} = - E_\mu^a \frac{\partial A(x,p)}{\partial x_\mu}, \qquad
    \left. \frac{\partial A}{\partial P^a} \right|_X = E_{a}^{\nu} \frac{\partial A}{\partial p^\nu} + \left. \frac{\partial x_\mu}{\partial P^a} \right|_X \frac{\partial A}{\partial x_\mu}.
\end{equation}
The remaining derivative is
\begin{equation}
    \left. \frac{\partial x_\mu}{\partial P^a} \right|_X =
    - X_b \frac{\partial E_\mu^b}{\partial P^a} = X_b \Gamma^b{}_{ac}E^c_\mu + \frac{X_a p_\mu}{m^2},
\end{equation}
where $\Gamma^b{}_{ac}$ is the Christoffel symbol of the induced metric. Substituting these relations into the canonical Poisson bracket, the terms involving $\Gamma^b{}_{ac}$ cancel under antisymmetrization, and we obtain
\begin{equation}
\{A(X_a, P^a), B(X_a, P^a)\}_{\rm P}
 = -\Pi_\mu{}^\nu
   \bigg[ \frac{\partial A}{\partial x_\mu} \frac{\partial B}{\partial p^\nu}
        - \frac{\partial A}{\partial p^\nu} \frac{\partial B}{\partial x_\mu} \bigg]
 + \frac{x_\mu p_\nu - x_\nu p_\mu}{m^2}
   \frac{\partial A}{\partial x_\mu} \frac{\partial B}{\partial x_\nu} .
\end{equation}
This is precisely the Dirac bracket associated with the mass-shell constraint and the gauge condition $x\cdot p=0$. Combining this result with the first-order expansion gives
\begin{equation}
    \{A, B\}_{\rm M}(x,p)
    = i \hbar \{A(x,p), B(x,p)\}_{\rm D}
    + {\cal O}(\hbar^2) .
\end{equation}
For the two-body phase space, the same construction is applied to each worldline before imposing the chosen two-body gauge conditions, yielding the Dirac bracket used in~\sec{sec:bracket}.

\bibliographystyle{JHEP}
\bibliography{references}
\end{document}